\documentclass[reprint,prb]{revtex4-2}
\usepackage{graphicx,amssymb,amsfonts,amsmath}
\usepackage{color}
\usepackage{mathtools}
\usepackage{latexsym}
\usepackage{graphicx}
\usepackage{float}
\usepackage{dcolumn}
\usepackage{bm}
\usepackage{amssymb}
\usepackage{amsmath}
\usepackage{mathtools}
\usepackage[space]{grffile}
\usepackage{xcolor}
\usepackage{comment}
\usepackage{hyperref}
\hypersetup{
    colorlinks=true,
    linkcolor=blue,
    filecolor=blue,      
    urlcolor=blue,
    citecolor=blue
    }

\begin{document}

\title{Floquet operator dynamics and orthogonal polynomials on the unit circle
}
\author{Hsiu-Chung Yeh}
\author{Aditi Mitra}
\affiliation{
Center for Quantum Phenomena, Department of Physics,
New York University, 726 Broadway, New York, New York, 10003, USA
}

\begin{abstract}
Operator spreading under stroboscopic time evolution due to a unitary is studied. An operator Krylov space is constructed and related to orthogonal polynomials on a unit circle (OPUC), as well as to the Krylov space of the edge operator of the Floquet transverse field Ising model with inhomogeneous couplings (ITFIM). The Verblunsky coefficients in the OPUC representation are related to the Krylov angles parameterizing the ITFIM. The relations between the OPUC and spectral functions are summarized and several  applications are presented. These include derivation of analytic expressions for the OPUC for persistent $m$-periodic dynamics, and the numerical construction of the OPUC for autocorrelations of the homogeneous Floquet-Ising model as well as the $Z_3$ clock model. The numerically obtained Krylov angles of the $Z_3$ clock model with long-lived period tripled autocorrelations show a spatial periodicity of six, and this observation is used to develop an analytically solvable model for the ITFIM that mimics this behavior.  
\end{abstract}
\maketitle
\section{Introduction}

Krylov subspace methods are a promising avenue for studying dynamics of complex systems as they typically map the problem in question to an effective single particle problem in one dimension, the Krylov chain \cite{Recbook,Schollwock05,SCHOLLWOCK201196}. For a hermitian operator undergoing continuous time evolution due to a Hamiltonian, this single particle problem is a nearest-neighbor hopping model, where the hopping parameters are in general inhomogeneous \cite{parker2019universal,Yates20a,Yates20,Yates22,nandy2024quantum,yeh2023slowly,rabinovici2021operator,rabinovici2022krylov,dymarsky2020quantum,dymarsky2021krylov}. For a hermitian operator undergoing stroboscopic dynamics due to a unitary, this single particle problem is the inhomogeneous Floquet transverse field Ising model (ITFIM) \cite{yeh2023universal,yeh2025moment}. In particular, the transverse field and Ising couplings are spatially inhomogeneous and represented as angles, dubbed the Krylov angles. 

The single-particle nature of the ITFIM is apparent via a Jordan-Wigner transformation that maps the ITFIM to a unitary that is bilinear in Majorana fermions. It should be noted that Hamiltonian dynamics too can be studied using an effective ITFIM if one studied the dynamics at stroboscopic times. This observation is helpful because often one may be able to make more precise statements about stroboscopic dynamics rather than continuous time dynamics. For example, Ref.~\cite{yeh2025moment} showed a direct relation between the Krylov angles and the stroboscopic autocorrelation such that the stroboscopic dynamics is constrained by unitarity. 

The recursion relations obeyed by the orthonormal basis of operators generated by Hamiltonian dynamics, are identical to the recursion relations obeyed by orthogonal polynomials \cite{Recbook,muck2022krylov}. This also generalizes to Floquet dynamics, where now the orthogonal polynomials lie on the unit circle \cite{szeg1939orthogonal,suchsland2023krylov,kolganov2025streamlinedkrylovconstructionclassification}. This connection is useful because analytic results for orthogonal polynomials can be used to capture non-trivial dynamics \cite{muck2022krylov}. Moreover, the dynamics can be studied directly using orthogonal polynomials without specifying underlying physical models. In particular, given a choice of OPUC, and an initial choice of an operator, an entire Krylov subspace can be generated.

Interestingly, the Floquet unitary in the Krylov basis has a simple 5-diagonal form when expressed in the Majorana representation of the ITFIM. When approaching the problem using OPUC, this 5-diagonal structure goes by the name of the CMV basis \cite{CANTERO200329,CANTERO200540,SIMON2007120}. 

The goal of this paper is to outline the connection between OPUC in general and the CMV basis in particular, and the physics approach of representing the Krylov subspace as a ITFIM. After making this connection, we derive analytical expressions for the OPUC for some simple, physically motivated problems. In addition, the $Z_3$ clock model is studied numerically, and an attempt is made to capture salient features of the dynamics of its long-lived edge mode by modeling it by an analytically solvable ITFIM. 

The paper is organized as follows. In Section \ref{OPUC} we express the stroboscopic dynamics of a general hermitian operator under a general unitary in terms of OPUC, showing how one may derive a convenient 5-diagonal basis known as the CMV basis. We then outline how the very same stroboscopic dynamics can be captured by the dynamics of an edge operator of the ITFIM. We then go on to explicitly relate the CMV and the ITFIM representations showing that the Verblusnky coefficients of the former are directly related to the Krylov angles of the latter. 
In Section \ref{Spectrum} we review relations between OPUC and spectral functions, and derive relations between the OPUC and the continued fraction expression for the Laplace transform of autocorrelation functions.  In Section \ref{App-exact} we derive analytic expressions for the OPUC for some simple $m$-periodic persistent dynamics where $m=1,2,3,4,6$. In Section \ref{AppZ3}, we numerically study long-lived and short-lived autocorrelations of the $Z_3$ clock model. The numerical results for the Krylov angles for the long-lived mode is then used to motivate a simple, analytically solvable model of an ITFIM that mimics the dynamics. This model is one where the Krylov angles have a spatial periodicity of six. Finally we conclude in Section \ref{Conc}. Eleven appendices provide intermediate details of the derivations as well as the OPUC for the Floquet transverse field Ising model with homogeneous transverse and Ising couplings. 

\section{OPUC in Floquet  dynamics} \label{OPUC}
The stroboscopic and unitary time evolution of an operator can be represented by a matrix in Krylov space that has an upper-Hessenberg form \eqref{Eq: upper-Hessenberg}. However, this representation is non-local and not sparse. A new basis introduced in Refs.~\cite{yeh2023universal,yeh2025moment} provides a sparse representation, which can be interpreted as the dynamics of the edge operator of an effective 1D Floquet inhomogeneous transverse-field Ising model (ITFIM). Interestingly, the same representation, known as the CMV basis \cite{simon2005orthogonalpart1}, can also be derived using orthogonal polynomials on the unit circle. Below, we outline in detail the connection between these two approaches.
\subsection{From Floquet Operator Krylov space to OPUC}
The unitary evolution of a Hermitian operator $O$ is described by $O(n) = (U_F^\dagger)^n O U_F^n$, where $U_F$ is the Floquet unitary and $n$ is the stroboscopic
time. By vectorization of  operators, the Floquet time evolution can be expressed as $|O(n)) = K^n |O)$, where $K$ is a superoperator defined as $K|O) = |U_F^\dagger O U_F)$. Therefore, the operator under Floquet dynamics expands a vector space into a Floquet operator Krylov space corresponding to $\{ |O), K|O), K^2|O), \ldots \}$. The inner product between two operator is defined as $(A|B) = \text{Tr}[A^\dagger B]/\text{Tr}[\mathbb{I}]$, where $\mathbb{I}$ is the identity operator. According to the Gram-Schmidt procedure, one constructs a set of orthonormal basis vectors $\{ |\mathcal{O}_k) \}$. There exists a linear transformation from $\{ |O(n)) \}$ to $\{ |\mathcal{O}_k) \}$\cite{Yates21,suchsland2023krylov,yeh2023universal}, 
\begin{align}
    |\mathcal{O}_k) = \sum_{n=0}^{k} \kappa_{kn}|O(n)), \label{Eq: linear trans two basis}
\end{align}
where $\kappa_{kn}$ encodes the details of the Gram-Schmidt procedure. This also implies
\begin{align}
   |{O}(k)) = \sum_{n=0}^{k} \kappa'_{kn}|\mathcal{O}_n), \label{Eq: linear trans two basis2}
\end{align}
such that $(\mathcal{O}_k|O(m)) = 0$ for $k > m$. We now follow the discussion in \cite{suchsland2023krylov}. By applying the definition $|O(n)) = K^n|O)$ and inserting the superoperator identity $\sum_{z} |z)(z|$ with $|z)$ being the 
eigenoperator of $K$, $K|z) = z|z)$, we obtain the following relation 
\begin{align}
    |\mathcal{O}_k)= \sum_{z} P_k(z)|z)(z|O),\  P_k(z) = \sum_{n=0}^{k} \kappa_{kn} z^n. \label{Eq: OPUC definition}
\end{align}
Since $K$ describes unitary dynamics, $z$ lies on the unit circle in the complex plane. Therefore, $\{ P_k(z) \}$ are called orthogonal polynomials on the unit circle (OPUC). 

The orthogonal conditions for OPUC can be derived from the orthogonality of the operator Krylov basis vectors. To show this, we start with \eqref{Eq: linear trans two basis} and consider the inner product $(\mathcal{O}_j|\mathcal{O}_k)$,
\begin{align}
    (\mathcal{O}_j|\mathcal{O}_k) = \sum_{n=0}^{j}\sum_{m=0}^{k} \overline{\kappa_{jn}} \kappa_{km} A(m-n),
    \label{Eq: O inner product}
\end{align}
where we have introduced the definition of  the autocorrelation function $A(n) = (O|K^n|O)$ and have applied the time translation property of the inner product $(O(n)|O(m)) = (O|O(m-n))$. 
Note that due to the cyclic property of the trace, the autocorrelation also obeys the property $A(n)=A(-n)$.

To proceed, we use the Fourier transformation relation between the autocorrelation $A(n)$ and the spectral function $\Phi(e^{i\omega})$ as below
\begin{align}
    &A(n) = \int_{-\pi}^\pi d\omega \Phi(e^{i\omega}) e^{i\omega n};\label{Eq: Fourier-1} \\&\Phi(e^{i\omega}) = \frac{1}{2\pi} \sum_{n=-\infty}^{\infty}A(n)e^{-i\omega n}.\label{Eq: Fourier-2}
\end{align}
Note that the spectral function is positive definite. To see this,  we insert the complete basis of the  Floquet unitary in \eqref{Eq: Fourier-2}, obtaining
\begin{align}
    \Phi(e^{i\omega}) = \frac{1}{\text{Tr}[\mathbb{I}]}\sum_{j,k} |\langle j|O|k\rangle|^2\delta_F(\omega-\epsilon_j+\epsilon_k),
    \label{Eq: spectral function}
\end{align}
where $U|k\rangle = e^{-i\epsilon_k}|k\rangle$, and $\delta_F(\omega)$ is the Floquet Dirac delta function: $\delta_F(\omega) = \sum_m\delta_D(\omega + 2\pi m)$ with $\delta_D(\omega)$ being the usual Dirac delta function. Therefore, $\Phi(e^{i\omega}) = \Phi(e^{-i\omega})$ and $\Phi(e^{i\omega}) \geq 0$.

By substituting $\eqref{Eq: Fourier-1}$ in \eqref{Eq: O inner product}, we obtain 
\begin{align}
    \delta_{jk} 
    &= \int_{-\pi}^{\pi}d\omega \Phi(e^{i\omega}) \overline{\left(\sum_{n=0}^{j} \kappa_{jn} e^{in\omega} \right)} \left(\sum_{m=0}^{k} \kappa_{km} e^{im\omega} \right)\nonumber\\
    &=\int_{-\pi}^{\pi}d\omega \Phi(e^{i\omega}) \overline{P_j(e^{i\omega})} P_k(e^{i\omega}),
\end{align}
where we have applied $(\mathcal{O}_k|\mathcal{O}_p) = \delta_{kp}$. Note that the spectral function serves as the weighting function of the inner product between the OPUC. With $z = e^{i\omega}$, we define the inner product of two generic polynomials, $F(z)$ and $H(z)$, on the unit circle as
\begin{align}
    \langle F(z),  H(z)\rangle \equiv \oint_{|z|=1} \frac{dz}{iz} \Phi(z)  \overline{F(z)} H(z) \label{Eq: polynomials inner product}.
\end{align}
Since $\Phi(e^{i\omega}) = \Phi(e^{-i\omega})$ and $\Phi(e^{i\omega}) \geq 0$, this definition of inner product of polynomials guarantees the following properties
\begin{align}
    &\overline{ \langle F(z),  H(z)\rangle} =  \langle H(z),  F(z)\rangle;\\
    &\langle F(z),  F(z)\rangle \geq 0.
\end{align}

Although we introduced OPUC from $\eqref{Eq: OPUC definition}$ with $|z| = 1$, in the following discussion, we will allow the OPUC to be evaluated on the whole complex plane. This is natural to do because  the poles insides the unit circle contribute to $\eqref{Eq: polynomials inner product}$ due to the residue theorem.

The above discussion shows the equivalence of inner product between operator Krylov space and OPUC:  $(\mathcal{O}_k|\mathcal{O}_p) = \langle P_k(z), P_p(z) \rangle$. Moreover, the dynamics in operator Krylov space can be represented as some equivalent computations in terms of OPUC. For example,
\begin{align}
    &(\mathcal{O}_k|K|\mathcal{O}_p) = \langle P_k(z),  zP_p(z)\rangle; \label{Eq: K-matrix in P}\\
    &(\mathcal{O}_k|O(m)) = \langle P_k(z), z^m \rangle.\label{Eq: P z^m inner prod}
\end{align}
The detailed derivation of the above results is presented in Appendix \ref{App: A}. Therefore, one can apply the techniques from OPUC to understand Floquet operator dynamics. Below, we demonstrate that OPUC serve as a powerful tool for computing the analytic expression for the matrix element $(\mathcal{O}_j|K|\mathcal{O}_k)$.

OPUC are known to satisfy the 
Szeg\H{o} recurrence relations \cite{simon2005orthogonalpart1} (see derivation in Appendix \ref{App: B})
\begin{subequations}
\label{Eq: Szego}
\begin{align}
    &\rho_kP_{k+1}(z) = zP_k(z) - \overline{\alpha_k}P_k^*(z);\label{Eq: Szego P-1}\\
    &\rho_k P_{k+1}^*(z) = P_k^*(z) - \alpha_k z P_k(z),\label{Eq: Szego P-2}
\end{align}
\end{subequations}
where $\alpha_k$ are the Verblunsky coefficients and $\rho_k$ is defined by $\rho_k = \sqrt{1-|\alpha_k|^2}$. The initial conditions are 
\begin{align}
P_0(z) = P_0^*(z) = 1.\label{initial}
\end{align}
Note that $P_k^*(z)$ is not the complex conjugation of $P_k(z)$ but called the $*$-reverse polynomials \cite{ismail2020encyclopedia} and defined as 
\begin{align}
    P_k^*(z) = z^k \overline{P_k(1/\overline{z})}. \label{Eq: P*}
\end{align}
With $\eqref{Eq: K-matrix in P}$ and $\eqref{Eq: Szego P-1}$, one has
\begin{align}
    (\mathcal{O}_j|K|\mathcal{O}_k)
    = \rho_k\langle P_j(z), P_{k+1}(z)\rangle + \overline{\alpha_k} \langle P_j(z), P_{k}^*(z)\rangle,
    \label{Eq: Krylov basis K-matrix}
\end{align}
where the first term is not zero when $j = k+1$. The second term is not zero when $0 \leq j \leq k$. This is because $P_k^*(z)$ is a polynomial of degree $k$ (namely, $P_k^*(z)$ consists of terms like $\{ 1, z, z^2, \ldots, z^k \}$) and $P_j(z)$ is orthogonal to any polynomial with degree smaller than $j$. This orthogonal condition comes from $\eqref{Eq: P z^m inner prod}$ and the Gram-Schmidt procedure $(\mathcal{O}_j|O(k)) = 0$ for $j > k$. With the analytic result of $\langle P_j(z), P_{k}^*(z)\rangle$ derived in Appendix \ref{App: C}, the matrix element of $(\mathcal{O}_j|K|\mathcal{O}_k)$ has the upper-Hessenberg form
\begin{align}
    &(\mathcal{O}_j|K|\mathcal{O}_k)=\left\{\begin{array}{cc}
     -\overline{\alpha_k}\alpha_{j-1} \prod_{l=j}^{k-1}\rho_l, & \text{for } 0\leq j \leq k ;\\
     \rho_k, & \text{for } j=k+1 ;\\
     0 & \text{else.}
     \end{array}\right.
     \label{Eq: upper-Hessenberg}
\end{align}
Although we obtain the analytic expression for $(\mathcal{O}_j|K|\mathcal{O}_k)$, the Floquet operator dynamics is non-local in Krylov orthonormal basis. As pointed out in Refs.~\cite{yeh2023universal,yeh2025moment}, one can construct a different basis, where Floquet dynamics of any Hermitian operator, generated by any unitary, can be mapped to the dynamics of the edge operator of an effective 1D Floquet inhomogeneous transverse-field Ising model (ITFIM). Moreover, the dynamics is local in this representation because the ITFIM is a local model comprising of single site and two-site unitaries. Interestingly, such a basis has been discovered by M.J. Cantero,  
L. Moral and L. Vel\'{a}zquez, now known as the CMV basis in the context of OPUC \cite{CANTERO200329,CANTERO200540,kolganov2025streamlinedkrylovconstructionclassification}. In the following subsection, we present the relation between the ITFIM and the CMV basis.

\subsection{ITFIM and the CMV basis}
\label{Sec: ITFIM and CMV basis}
We first summarize the results of the mapping of the Krylov dynamics to the ITFIM from Ref. \cite{yeh2023universal,yeh2025moment}. The 1D Floquet ITFIM parametrized by Krylov angles, and with open boundary conditions is
\begin{subequations}
\label{Eq: ITFIM}
\begin{align}
&U_{F} = U_z U_{xx};\\
&U_z = \prod_{l=1}^{N/2} e^{-i\frac{\theta_{2l-1}}{2}\sigma_l^z} ;
\ U_{xx} = \prod_{l=1}^{(N-2)/2} e^{-i\frac{\theta_{2l}}{2}\sigma_l^x \sigma_{l+1}^x}.
\end{align}  
\end{subequations}
Above, $\sigma^{x,z}_l$ are the Pauli  matrices on site $l$, we assume $N$ is even  and $\{ \theta_1, \ldots, \theta_{N-1} \} \in [0,\pi]$ are the Krylov angles. The odd Krylov angles $\theta_{2l-1}$ denote the strength of the transverse field on site $l$ while the even Krylov angles $\theta_{2l}$ denote the strength of the Ising interaction between the spins on sites $l,l+1$. 
The above model is effectively a non-interacting model. To see this, we define the Jordan-Wigner transformation
\begin{align}
  &\gamma_{2\ell -1} = \prod_{j=1}^{\ell-1}\sigma^z_j \sigma^x_\ell; &\gamma_{2\ell} = \prod_{j=1}^{\ell-1}\sigma^z_j \sigma^y_\ell.
  \label{Eq: Jordan-Wigner}
\end{align}
After a Jordan-Wigner transformation, $U_F$ is bilinear in the Majorana fermions since
\begin{align}
U_z = \prod_{l=1}^{N/2} e^{-\frac{\theta_{2l-1}}{2}\gamma_{2l-1}\gamma_{2l}} ;
U_{xx} = \prod_{l=1}^{(N-2)/2} e^{-\frac{\theta_{2l}}{2}\gamma_{2l}\gamma_{2l+1}}.
\end{align}

Let us consider an operator $\Psi$ that can be written as a linear combination of Majoranas, $\Psi = \sum_{k}\psi_k\gamma_k$. The coefficients can be viewed as a column vector $\vec{\psi} = (\psi_1, \psi_2, \psi_3, \ldots)^\intercal$. Under the effect of the two unitaries, the coefficients transform according to, $U_z^\dagger \Psi U_z = \sum_{k} \psi_k' \gamma_k$ with $\vec{\psi}' = M_z \vec{\psi}$  and $U_{xx}^\dagger \Psi U_{xx} = \sum_{k} \psi_k'' \gamma_k$ with $\vec{\psi}'' = M_{xx} \vec{\psi}$. For example for $N=6$
\begin{widetext}
\begin{align}
&M_z= \begin{pmatrix}
         \cos\theta_1 & \sin\theta_1 & 0 & 0 & 0 & 0 \\
 -\sin\theta_1 & \cos \theta_1 & 0 & 0 & 0 & 0 \\
 0 & 0 & \cos \theta_3 & \sin \theta_3 & 0 & 0 \\
 0 & 0 & -\sin \theta_3 & \cos \theta_3 & 0 & 0 \\
 0 & 0 & 0 & 0 & \cos \theta_5 & \sin \theta_5 \\
 0 & 0 & 0 & 0 & -\sin \theta_5 & \cos \theta_5
\end{pmatrix};
&M_{xx}=
\begin{pmatrix}
    1 & 0 & 0 & 0 & 0 & 0 \\
 0 & \cos \theta_2 & \sin \theta_2 & 0 & 0 & 0 \\
 0 & -\sin \theta_2 & \cos \theta_2 & 0 & 0 & 0 \\
 0 & 0 & 0 & \cos \theta_4 & \sin \theta_4 & 0 \\
 0 & 0 & 0 & -\sin \theta_4 & \cos \theta_4 & 0 \\
 0 & 0 & 0 & 0 & 0 & 1
\end{pmatrix}.
\end{align}
Denoting the unitary evolution in the Majorana basis after one time step as
\begin{align}
   (\gamma_j|K_I|\gamma_k) = \frac{1}{\text{Tr}[\mathbb{I}]}{\rm Tr} \biggl[\gamma_j U_F^{\dagger}\gamma_kU_F\biggr], 
\end{align}
$K_I$ in the single Majorana basis has the form
\begin{align}
    K_I &=  M_{xx} M_z 
    = \begin{pmatrix}
        \cos \theta_1 & \sin \theta_1 & 0 & 0 & 0 & 0 \\
 -\sin \theta_1 \cos \theta_2 & \cos
   \theta_1 \cos \theta_2 & \sin \theta_2
   \cos \theta_3 & \sin \theta_2 \sin \theta_3 & 0 & 0 \\
 \sin \theta_1 \sin \theta_2 & -\cos \theta_1\sin \theta_2 & \cos \theta_2 \cos
   \theta_3 & \cos \theta_2\sin \theta_3  & 0
   & 0 \\
 0 & 0 & -\sin \theta_3 \cos \theta_4 & \cos
   \theta_3 \cos \theta_4 & \sin \theta_4
   \cos \theta_5 & \sin \theta_4 \sin \theta_5 \\
 0 & 0 & \sin \theta_3 \sin \theta_4 & -\cos \theta_3\sin \theta_4  & \cos \theta_4
   \cos \theta_5 & \cos \theta_4 \sin \theta_5  \\
 0 & 0 & 0 & 0 & -\sin \theta_5 & \cos \theta_5
\end{pmatrix}.
\label{Eq: K matrix}
\end{align}
\end{widetext}
Generalizing the above result to arbitrary system sizes, the matrix elements in the Majorana basis of the ITFIM have the following form
\begin{subequations}
\label{Eq: Majorana basis matrix element}
\begin{align}
    &(\gamma_{2k+1}|K_I|\gamma_{2k-1}) = \sin\theta_{2k-1}\sin\theta_{2k};\\ &(\gamma_{2k+1}|K_I|\gamma_{2k}) = -\cos\theta_{2k-1}\sin\theta_{2k};\\
    &(\gamma_{2k+1}|K_I|\gamma_{2k+1}) = \cos\theta_{2k}\cos\theta_{2k+1};\\ 
    &(\gamma_{2k+1}|K_I|\gamma_{2k+2}) = \cos\theta_{2k}\sin\theta_{2k+1};\\
    &(\gamma_{2k+2}|K_I|\gamma_{2k+1}) = -\sin\theta_{2k+1}\cos\theta_{2k+2};\\ 
    &(\gamma_{2k+2}|K_I|\gamma_{2k+2}) = \cos\theta_{2k+1}\cos\theta_{2k+2};\\
    &(\gamma_{2k+2}|K_I|\gamma_{2k+3}) = \sin\theta_{2k+2}\cos\theta_{2k+3};\\ 
    &(\gamma_{2k+2}|K_I|\gamma_{2k+4}) = \sin\theta_{2k+2}\sin\theta_{2k+3},
\end{align}
\end{subequations}
where the boundary condition $\theta_0 = \theta_N = 0$ is applied. Refs. \cite{yeh2023universal,yeh2025moment} derived exact relations between the matrix elements of the $N$-dimensional unitary in its upper-Hessenberg form \eqref{Eq: upper-Hessenberg} and the first $N$ Krylov angles of the ITFIM.  Therefore, the ITFIM can reproduce any Floquet dynamics of Hermitian operators. In other words, the initial operator $O$ in the original model can be identified as $\gamma_1$ of the ITFIM. However, this equivalence holds only at the level of the autocorrelation. They remain fundamentally different at the operator level, i.e., in their algebraic structure.

We now present the CMV basis \cite{simon2005orthogonalpart1} and its relation to the Majorana representation of  the ITFIM. The CMV basis is related to the OPUC as follows
\begin{align}
    &\chi_{2k} = z^{-k}P_{2k}; 
    &\chi_{2k+1} = z^{-k-1}P_{2k+1}^*.
    \label{Eq: CMV basis}
\end{align}
Now we show how \eqref{Eq: Majorana basis matrix element} is recovered through computations in the CMV basis.  For this, the following identities are helpful (see Appendix \ref{App: D} for details),
\begin{subequations}
\label{Eq: CMV basis matrix element}
\begin{align}
    &\langle \chi_{2k}, z\chi_{2k-2} \rangle = \rho_{2k-2}\rho_{2k-1},\\
    &\langle \chi_{2k}, z\chi_{2k-1} \rangle = -\overline{\alpha_{2k-2}}\rho_{2k-1},\\  
    &\langle \chi_{2k}, z\chi_{2k} \rangle = -\alpha_{2k-1}\overline{\alpha_{2k}},\\
    &\langle \chi_{2k}, z\chi_{2k+1} \rangle = -\alpha_{2k-1}\rho_{2k},\\
    &\langle \chi_{2k+1}, z\chi_{2k} \rangle = \rho_{2k}\overline{\alpha_{2k+1}},\\  
    &\langle \chi_{2k+1}, z\chi_{2k+1} \rangle = -\alpha_{2k}\overline{\alpha_{2k+1}},\\  
    &\langle \chi_{2k+1}, z\chi_{2k+2} \rangle = \rho_{2k+1}\overline{\alpha_{2k+2}},\\
    &\langle \chi_{2k+1}, z\chi_{2k+3} \rangle = \rho_{2k+1}\rho_{2k+2},
\end{align}
\end{subequations}
where all other matrix elements are zero. For the unitary evolution of Hermitian operators, the matrix element in \eqref{Eq: upper-Hessenberg} is real \cite{yeh2023universal}. By comparison between \eqref{Eq: Majorana basis matrix element} and \eqref{Eq: CMV basis matrix element}, we have the following relations between the matrix elements in the two representations 
\begin{align}
&(\gamma_{j+1}|K_I|\gamma_{k+1}) =\langle \chi_j, z\chi_k \rangle, \label{eq: K-matrix-cmv}\\
    &\alpha_k = (-1)^{k}\cos\theta_{k+1},\ \rho_k = \sin\theta_{k+1}.\label{Eq: Verblunsky Krylov angle}
\end{align}
Thus the Verblunsky coefficients  $\alpha_k$ (these can be taken to be real) and the Krylov angles $\theta_{k+1}$ are directly related.
To summarize, if the matrix elements of the unitary generating operator time-evolution are expressed in the basis of the  OPUC (as in \eqref{Eq: K-matrix in P}), one recovers the upper-Hessenberg form \eqref{Eq: upper-Hessenberg}. In contrast, if the matrix elements of the unitary are expressed in the CMV basis \eqref{eq: K-matrix-cmv}, one recovers  the representation in the Majorana basis of the ITFIM, \eqref{Eq: Majorana basis matrix element}.

In this work, we will focus on $\alpha_k$ being real as we are interested in the Floquet unitary dynamics of Hermitian operators. If one considers the Floquet unitary dynamics of wave functions, their corresponding Krylov space is also related to the OPUC in the CMV basis, however the Verblunsky coefficients can be complex, see discussion in \cite{kolganov2025streamlinedkrylovconstructionclassification}.

\section{OPUC as an approximate spectral function}\label{Spectrum}
In this section, we introduce the Bernstein–Szeg\H{o} approximation, which is useful for constructing approximate spectral functions in numerical computations. Moreover, we relate this approximation to the continued fraction expansion in the Laplace transform \cite{yeh2025moment} and show that the Szeg\H{o} recurrence relations \eqref{Eq: Szego} are embedded within the recurrence relations of the continued fraction. 
\subsection{Bernstein–Szeg\H{o} approximation} \label{Sec: Bernstein approx}
We have presented the relation between operators in Krylov space and OPUC. In addition, the OPUC also contain information about the spectral function, and can be used to approximate it. The Bernstein–Szeg\H{o} approximation for the spectral function is \cite{simon2005orthogonalpart1}
\begin{align}
    \Phi_k(e^{i\omega}) = \frac{1}{2\pi}   \frac{1}{|P_k(e^{i\omega})|^2}.
    \label{Eq: Bernstein approx}
\end{align}
The exact spectral function is recovered by taking the following limit
\begin{align}
    \Phi(e^{i\omega}) = \frac{1}{2\pi} \lim_{k\rightarrow \infty}  \frac{1}{|P_k(e^{i\omega})|^2}.
    \label{Eq: Bernstein large k}
\end{align}
The approximate $\Phi_k(e^{i\omega})$ shares the same Verblunsky coefficient $\{ \alpha_0,\alpha_1, \ldots, \alpha_{k-1} \}$ as $\Phi(e^{i\omega})$ but $\alpha_{j\geq k} = 0$. For real $\alpha_k$, this implies that the Krylov angles $\theta_{j\geq k+1} = \pi/2$, namely the bulk of the ITFIM becomes a dual unitary model and is maximally ergodic \cite{suchsland2023krylov}. Before proving the above properties of the Bernstein-Szeg\H{o} approximation, we have to first prove a key property that the OPUC has zeros only for $|z| < 1$. Suppose that $P_k(z)$ has a zero at $z_0$, we can express $P_k(z)$ as
\begin{align}
    P_k(z) = (z-z_0)\tilde{P}_k(z),
\end{align}
where $\tilde{P}_k(z)$ is a polynomial with degree $k-1$. Therefore,
\begin{align}
    \langle z\tilde{P}_k(z), z\tilde{P}_k(z)\rangle = \langle P_k(z) + z_0\tilde{P}_k(z), P_k(z) + z_0\tilde{P}_k(z) \rangle.
\end{align}
On the left hand side, we have $\langle z\tilde{P}_k(z), z\tilde{P}_k(z)\rangle = \langle \tilde{P}_k(z), \tilde{P}_k(z)\rangle$ due to the definition of the inner product \eqref{Eq: polynomials inner product}. Note that the inner product is defined on the unit circle and therefore one may apply $\overline{z} = 1/z$ in \eqref{Eq: polynomials inner product}. On the right hand side, there will be no cross terms like $\langle P_k(z), \tilde{P}_k(z)\rangle = 0$ since $P_k(z)$ is orthogonal to polynomials with degree lower than $k$. Therefore, we arrive at the following relation
\begin{align}
    \langle P_k(z) , P_k(z) \rangle = (1-|z_0|^2)\langle \tilde{P}_k(z), \tilde{P}_k(z)\rangle > 0
\end{align}
where the norm of the polynomials are non-negative. In addition, since $\langle P_k(z) , P_k(z) \rangle = 1$, it follows that $|z_0| < 1$. From this, we can conclude that $P_k(z)$ has zeros only for $|z|<1$. In addition, $P_k^*(z)$ has zeros for $|z| > 1$ according to the definition $\eqref{Eq: P*}$.

Below, we show that $P_k(z)$ is the OPUC corresponding to the approximate spectral function $\Phi_k(e^{i\omega})$. First, we show that $P_k(z)$ is normalized
\begin{align}
    \langle P_k(z), P_k(z) \rangle_{\Phi_k} &= \oint_{|z|=1} \frac{dz}{i z} \Phi_k(z) |P_k(z)|^2\nonumber\\
    &=\oint_{|z|=1} \frac{dz}{2\pi i z} = 1,
\end{align}
where the subscript $\Phi_k$ indicates replacing $\Phi(z)$ by $\Phi_k(z)$ in \eqref{Eq: polynomials inner product}. Note that one can apply $e^{i\omega} = z$ in the inner product computation.
Second, we show that $P_k(z)$ is orthogonal to $z^l$ for $0 \leq l < k$.
\begin{align}
    \langle z^l, P_k(z)\rangle_{\Phi_k} = \oint_{|z|=1} \frac{dz}{iz} \frac{\overline{z^l}P_k(z)}{2\pi |P_k(z)|^2}.
\end{align}
Then we use $\overline{z} = 1/z$ and apply the *-reverse relation \eqref{Eq: P*} inside the inner product to obtain  
\begin{align}
    \langle z^l, P_k(z)\rangle_{\Phi_k}
    = \oint_{|z|=1} \frac{dz}{2\pi i} \frac{z^{k-l-1}}{P_k^*(z)} = 0,\ \forall 0 \leq l \leq k-1
\end{align}
where we have applied the residue theorem and the fact that there are no poles for $|z| < 1$ for $0 \leq l \leq k-1$. Therefore, $P_k(z)$ is the OPUC with degree $k$ corresponding to the spectral function  $\Phi_k(\omega)$. Note that $\{ P_0(z), P_1(z), P_2(z), \ldots, P_{k-1}(z) \}$ are uniquely determined once $P_k(z)$ is fixed. From the Szeg\H{o} recurrence relations, the OPUC are constructed iteratively with a fixed set of Verblunsky coefficients and this correspondence is unique. Therefore, fixing $P_k(z)$ uniquely determines $\{ \alpha_0, \alpha_1, \ldots, \alpha_{k-1} \}$ and all the previous OPUC $\{ P_0(z), P_1(z), P_2(z), \ldots, P_{k-1}(z) \}$, see detailed discussion in Appendix \ref{App: E}.

Finally, we will show that $z^l P_k(z)$ is the $l+k$ order OPUC with the spectral function $\Phi_k(e^{i\omega})$ for $l \geq 0$.  Let us first check the normalization condition
\begin{align}
    \langle z^lP_k(z), z^lP_k(z) \rangle_{\Phi_k} = \langle P_k(z), P_k(z) \rangle_{\Phi_k} = 1,
\end{align}
where $\overline{z} = 1/z$ is applied. Next, we consider the orthogonal condition
\begin{align}
    \langle z^j, z^lP_k(z) \rangle_{\Phi_k}
    &= \oint_{|z|=1} \frac{dz}{2\pi i} \frac{z^{l+k-j-1}}{P_k^*(z)}\nonumber\\
    &= 0,\ \forall\ 0 \leq j \leq l+k-1
\end{align}
Therefore, $z^lP_k(z)$ is orthogonal to any polynomial with degree lower than $l+k$ and it is an $l+k$ order OPUC with $\Phi_k(e^{i\omega})$.

In summary, we have shown that the Bernstein-Szeg\H{o} approximation for the spectral function  $\Phi_k(e^{i\omega})$ corresponds to the following set of OPUC, $\{ P_0(z), P_1(z), \ldots, P_k(z), zP_k(z), z^2P_k(z),\ldots \}$. Applying  \eqref{Eq: Szego P-1}, one concludes $\alpha_{j\geq k} = 0$. The true spectral function can be recovered by taking the large $k$ limit. Truncating at suitably large $k$ is a good practical approximation as many non-integrable systems show a tendency towards vanishing Verblunsky coefficients in numerical studies \cite{suchsland2023krylov,yeh2023universal}. At the dual unitary point $\theta_k=\pi/2$, the OPUC take the particularly simple form $P_k(z)= z^k$ according to \eqref{Eq: Szego}. In addition, the corresponding spectral function is uniform \cite{suchsland2023krylov}.

A special case of the Bernstein–Szeg\H{o} approximation arises when the Krylov space is finite. In this case, only a finite number of Verblunsky coefficients $\{\alpha_0, \alpha_1, \ldots, \alpha_{k^*}  \}$ exist, with $|\alpha_{k^*}| = 1$ (equivalently, $\theta_{k^*+1} = 0$ or $\pi$). In this situation, the Szeg\H{o} recurrence relation \eqref{Eq: Szego P-1} fails to generate $P_{k^*+1}$ since $\rho_{k^*} = 0$ when $|\alpha_k^*| = 1$. Since the Krylov dimension is finite, there are only a finite number of eigenvalues, and  therefore the spectral function consists of a finite number of delta-function peaks. Nevertheless, the Bernstein–Szeg\H{o} approximation remains valid in the following sense. Without loss of generality, we take $\alpha_{k^*} = 1$, in which case the spectral function for this finite Krylov space can be recovered via the limit
    \begin{align}
        \Phi(e^{i\omega}) = \frac{1}{2\pi} \lim_{\epsilon \rightarrow 0^+} \frac{1}{|P_{k^*+1}(e^{i\omega})|^2}\Big|_{\alpha_{k^*}=e^{-\epsilon}}.
    \end{align}
    For any finite  $\epsilon$, $P_{k^*+1}$ is regularized and the true spectral function is obtained in the limit $\epsilon \rightarrow 0^+$. We present the example for $\alpha_0 = 1$ in Appendix \ref{App: F}.
    
Finally, we conclude this subsection with Rakhmanov’s theorem \cite{simon2005orthogonalpart1}. A spectral function can be decomposed into smooth and singular components. If the smooth part is positive almost everywhere (exceptions are measure zero), then the Verblunsky coefficients must vanish asymptotically, $\lim_{k\rightarrow\infty} \alpha_k = 0$, or equivalently $\lim_{k\rightarrow\infty} \theta_k = \pi/2$. Physically, this condition corresponds to a gapless spectral function, since a gapped spectrum would require the function to vanish over an interval of frequencies, which is not a set of measure zero.

\subsection{Laplace transformation and its continued fraction expansion}
Another interesting property of the spectral function is its relation to the Laplace transformation. For a given autocorrelation, we can define its discrete Laplace transformation $G_L(z)$ as
\begin{align}
    G_L(z) = \sum_{n=0}^{\infty} A(n)z^{-n},
    \label{Eq: G Laplace}
\end{align}
where $z$ is complex number with $|z|>1$ ensuring a convergent expression for $G_L(z)$. According to \eqref{Eq: Fourier-2},  the spectral function $\Phi(e^{i\omega})$ is related to $G_L(z)$ as follow
\begin{align}
    \Phi(e^{i\omega}) = \frac{1}{2\pi}\left(-1 + 2\lim_{\epsilon\rightarrow 0^+} \text{Re}[G_L(e^{\epsilon + i\omega})]\right),
    \label{Eq: spectral Laplace}
\end{align}
where the $-1$ in the round bracket accounts for the double counting of $A(0)$ where $A(0) = 1$. For ITFIM, $G_L(z)$ has an equivalent representation in terms of continued fractions parametrized by Krylov angles \cite{yeh2025moment}. Below we omit the $z$-dependence for compactness.
\begin{subequations}
\label{Eq: Laplace Continued fraction}
\begin{align}
    &G_L(z,\theta) = \underset{M\rightarrow \infty}{\lim}G_C(z,\theta; M), \forall\ |z|>1;\\
    &G_C(z,\theta;M) = \frac{a_0}{b_0 + \frac{a_1}{  \ddots+\frac{a_M}{b_M}}} =\frac{f_M}{h_M},
\end{align}
\end{subequations}
where
\begin{subequations}\label{Eq: a b coefficient}
\begin{align}
    &a_0 = z;\quad\quad b_0 = z - \cos\theta_1; \\
    &(a_{k}, b_{k}) = \Big\{\begin{array}{cc}
    (\sin^2\theta_k,\ z\cos\theta_{k+1} - \cos\theta_k) & \text{if}\ k\ \text{is odd};\\
     (z^2\sin^2\theta_k,\ z\cos\theta_{k} - \cos\theta_{k+1}) & \text{else}.
    \end{array}
\end{align}
\end{subequations}
$f_M$ and $h_M$ are polynomials obeying the following recurrence relations,
\begin{align}
    &f_k = b_k f_{k-1} + a_k f_{k-2};\ h_k = b_k h_{k-1} + a_k h_{k-2},
    \label{Eq: f h recurrence}
\end{align}
with initial condition $f_{-2} = 1, f_{-1} = 0, h_{-2} = 0$ and $h_{-1} = 1$. 

From \eqref{Eq: spectral Laplace} and \eqref{Eq: Laplace Continued fraction}, one may expect that the truncated continued fraction is related to the OPUC via the Bernstein-Szeg\H{o} approximation \eqref{Eq: Bernstein approx}. In fact, they obey the following relations (see details in Appendix \ref{App: G})
\begin{subequations}
\label{Eq: Bernstein approx truncated continued fraction}
\begin{align}
    &\frac{1}{|P_{2k}(e^{i\omega})|^2} \nonumber\\
    &= -1 + 2\lim_{\epsilon\rightarrow 0^+} \text{Re}[G_C(e^{\epsilon + i\omega},\theta;2k)]\Big|_{\theta_{2k+1}=\frac{\pi}{2}};\label{Eq: P GC-1}\\
    &\frac{1}{|P_{2k+1}(e^{i\omega})|^2}\nonumber\\
    &= -1 + 2\lim_{\epsilon\rightarrow 0^+} \text{Re}[G_C(e^{\epsilon + i\omega},\theta;2k)].\label{Eq: P GC-2}
\end{align}
\end{subequations}
Note that $P_{2k}$ depends on $\{ \theta_1, \ldots \theta_{2k} \}$ and $G_C(z,\theta;2k)$ depends on $\{ \theta_1, \ldots \theta_{2k+1} \}$ according to \eqref{Eq: a b coefficient}. Therefore, in \eqref{Eq: P GC-1}, we have to set $\theta_{2k+1} = \pi/2$ for $G_C(z,\theta;2k)$ to match the number of degrees of freedom. \eqref{Eq: Bernstein approx truncated continued fraction} suggests that the Szeg\H{o} recurrence relations \eqref{Eq: Szego} are hidden in the recurrence relation of the continued fraction \eqref{Eq: f h recurrence}. Appendix \ref{App: G} shows how the Szeg\H{o} recurrence relations are implied by the continued fraction recurrence relations \eqref{Eq: f h recurrence}.

\section{OPUC for persistent $m$-period autocorrelation functions}\label{App-exact}
\label{Sec: m-period}
We consider persistent $m$-period autocorrelations of the form
\begin{align}
A^{(m)}(n>0) = A\cos(2\pi n/m), \,\, A^{(m)}(n=0)=1. \label{Eq: A m-period}
\end{align}
where $A$ is a real number within $[0,1]$ and $m$ is an integer. Note that, for $A = 1$, the Gram-Schmidt procedure stops after a finite number of steps because for this case $A^{(m)}(m) = (O(m)|O)=1$ and $A^{(2m)}(m) = (O(m)|O)=-1$.  In particular, the Krylov space dimension of $A^{(m)}$ is finite with dimension  $m$ for $ m\in$ odd and $m/2$ for $m\in$ even, where for the latter the minus-sign can be absorbed in the Gram-Schmidt procedure. In the following discussion, we focus on $A \neq 1$ and the Krylov space dimension is therefore infinite.

Below we summarize the discrete Laplace transformation and the corresponding Krylov angles for $m = 1, 2, 3, 4, 6$ \cite{yeh2025moment}. Besides the OPUC for the $m=1$ case \cite{simon2005orthogonalpart1}, to the best of our knowledge, the OPUC for the other cases have not been explicitly reported before.

We first review the results for $m = 1, 2, 4$ as they are related to each other. For $m = 1$, $G_L^{(1)}(z)$ is
\begin{align}
    G_L^{(1)}(z) = \frac{z-1+A}{z-1}.\label{Eq: GL1}
\end{align}
The above can be written in the continued fraction representation $G_C^{(1)}(z,\theta^{(1)},M)$ with the following Krylov angles
\begin{align}
    \cos\theta^{(1)}_k = \frac{(-1)^{k-1}A}{1+(k-1)A}.\label{Eq: m=1 Krylov angle}
\end{align}
From the Szeg\H{o} recurrence relations \eqref{Eq: Szego}, the OPUC can be deduced from the Krylov angles,
\begin{align}
    &P^{(1)}_k(z) \nonumber\\
    &= \left(\prod_{j=1}^{k} \frac{1}{\sin\theta^{(1)}_j} \right)\left[z^k - (-1)^{k-1}\cos\theta^{(1)}_k\sum_{j=0}^{k-1}z^j\right].
    \label{Eq: P m=1}
\end{align}
See details in Appendix \ref{App: H} where other cases are also discussed.  The spectral function can be derived from \eqref{Eq: GL1} and \eqref{Eq: spectral Laplace}
\begin{align}
    \Phi^{(1)}(e^{i\omega}) = \frac{1-A}{2\pi} + A \delta_F(\omega),
    \label{Eq: spectral function m=1}
\end{align}
where we applied the Floquet Sokhotski–Plemelj theorem, see Appendix \ref{App: I} for its application to $m=1$ and other cases.

For $m = 2$, $G_L^{(2)}(z)$ is
\begin{align}
    G_L^{(2)}(z) = \frac{-z-1+A}{-z-1}.
\end{align}
It is related to the $m = 1$ case by reflection, $G_L^{(2)}(z) = G_L^{(1)}(-z)$. The reflection in $z$ leaves the even Krylov angles unchanged, while transforming the odd Krylov angles to their supplementary angles \cite{yeh2025moment}. Therefore, $G_C^{(2)}(z,\theta^{(2)},M)$ is parametrized by
\begin{align}
    \theta^{(2)}_k = \bigg\{\begin{array}{cc}
    \pi - \theta^{(1)}_k & \text{if}\ k\ \text{is odd};\\
     \theta^{(1)}_k & \text{else}.
    \end{array}
    \label{Eq: m=2 Krylov angle}
\end{align}
The reflection property implies the following relations for the OPUC (see Appendix \ref{App: H})
\begin{align}
    &P^{(2)}_k(z) = (-1)^k P^{(1)}_k(-z),\ P^{(2)*}_k(z) = P^{(1)*}_k(-z).
    \label{Eq: P m=2}
\end{align}
The corresponding spectral functions are related as
\begin{align}
    \Phi^{(2)}(e^{i\omega}) = \Phi^{(1)}(e^{i(\omega-\pi)})
\end{align}

For $m = 4$, $G_L^{(4)}(z)$ is given by
\begin{align}
    G_L^{(4)}(z) = \frac{-z^2-1+A}{-z^2-1}.
\end{align}
It is invariant under reflection, $G_L^{(4)}(z) = G_L^{(4)}(-z)$, and related to $m = 1$ by square-reflection, $G_L^{(4)}(z) = G_L^{(1)}(-z^2)$. Due to the reflection invariance, all odd Krylov angles of $G_C^{(4)}(z,\theta^{(4)},M)$ are $\pi/2$ as $\pi/2$ is its own supplementary angle. The even angles of $\theta^{(4)}$ are mapped one-to-one to $\theta^{(1)}$ according to square-reflection \cite{yeh2025moment}. In summary the Krylov angles of  $G_C^{(4)}(z,\theta^{(4)},M)$ obey
\begin{align}
    \theta^{(4)}_k = \bigg\{\begin{array}{cc}
    \pi/2 & \text{if}\ k\ \text{is odd};\\
     \theta^{(1)}_{k/2} & \text{else}.
    \end{array}.
    \label{Eq: m=4 Krylov angle}
\end{align}
The corresponding OPUC obey (see Appendix \ref{App: H})
\begin{subequations}
\label{Eq: P m=4}
\begin{align}
    &P^{(4)}_{2k}(z) = (-1)^k P^{(1)}_k(-z^2),\ P^{(4)*}_{2k}(z) = P^{(1)*}_k(-z^2);\\ 
    &P^{(4)}_{2k+1}(z) = zP^{(4)}_{2k}(z),\ P^{(4)*}_{2k+1}(z) = P^{(4)*}_{2k}(z),
\end{align}
\end{subequations}
where the second equation is simply a consequence of the Szeg\H{o} recurrence relations \eqref{Eq: Szego} with $\alpha_{2k} = (-1)^{2k}\cos\theta^{(4)}_{2k+1} = 0$ and $\rho_{2k} = \sin\theta^{(4)}_{2k+1} = 1$.
The corresponding spectral function is (see Appendix \ref{App: J})
\begin{align}
    \Phi^{(4)}(e^{i\omega}) = \frac{1}{2}\left[ \Phi^{(1)}(e^{i(\omega-\pi/2)}) +  \Phi^{(1)}(e^{i(\omega+\pi/2)})\right].
\end{align}

\begin{figure*}
    \includegraphics[width=0.32\textwidth]{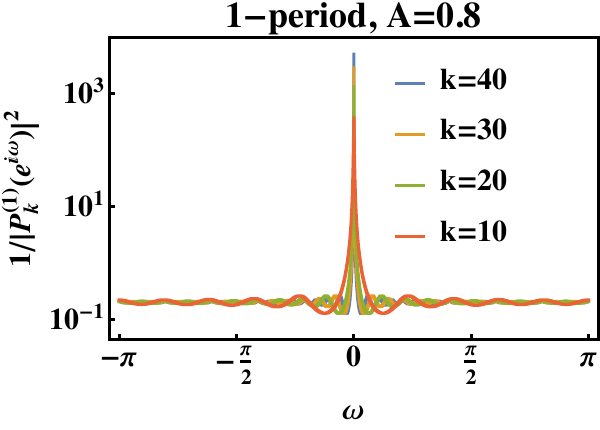}
    \includegraphics[width=0.32\textwidth]{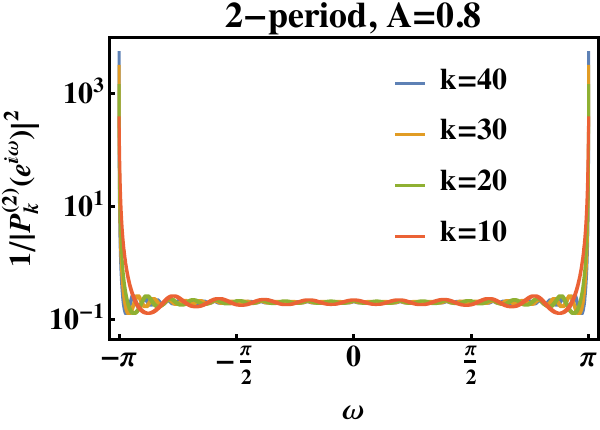}
    \includegraphics[width=0.32\textwidth]{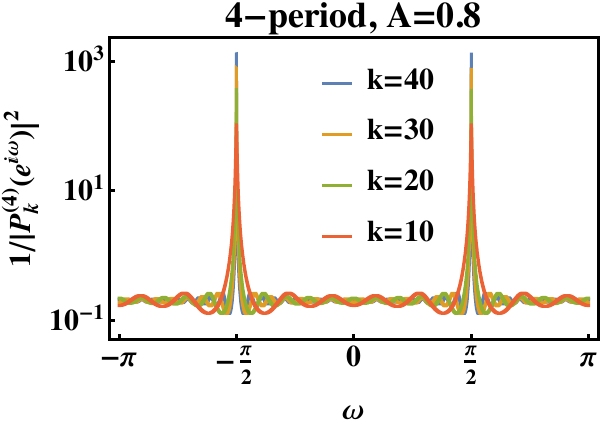}
    \includegraphics[width=0.32\textwidth]{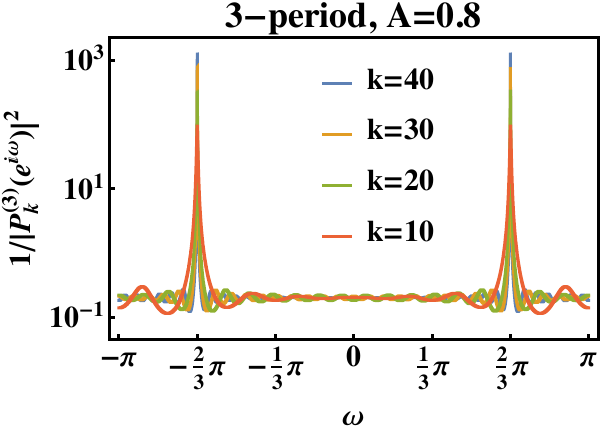}
    \includegraphics[width=0.32\textwidth]{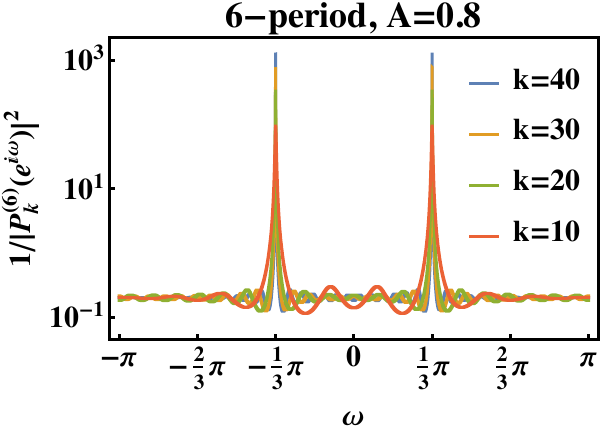}    

    \caption{The numerical results for $1/|P^{(m)}_k(e^{i\omega})|^2$ for $m = 1,2,4$ (\eqref{Eq: P m=1}, \eqref{Eq: P m=2}, \eqref{Eq: P m=4}) in the top panels and $m = 3,6$ (\eqref{Eq: P m=3}, \eqref{Eq: P m=6}) in the bottom panels with $k = 10,20,30,40$ and $A=0.8$. As $k$ increase, the peaks become sharper indicating the delta function behavior of the spectral function. This agrees with the Bernstein–Szeg\H{o} approximation \eqref{Eq: Bernstein large k}. Besides the delta function, the background approaches $(1-A)$ as expected from \eqref{Eq: spectral function m=1}.}  
    \label{Fig: m-period OPUC}
\end{figure*}

We now review the $m= 3, 6$ cases. They obey similar relations as the $m = 1, 2$ cases. For $m = 3$, the discrete Laplace transform  $G_L^{(3)}(z)$ is 
\begin{align}
    G_L^{(3)}(z) = \frac{1-A + \left(1-\frac{A}{2}\right)z+z^2}{1+z+z^2}.
\end{align}
\begin{widetext}
The corresponding Krylov angles of the continued fraction representation $G_C^{(3)}(z,\theta^{(3)},M)$ are
\begin{align}
\label{Eq: m=3 Krylov angle}
    \cos\theta^{(3)}_{3k-2} = \frac{(-1)^kA}{2+3(k-1)A},\quad\quad
    \cos\theta^{(3)}_{3k-1} = \frac{(-1)^{k-1}A}{2+3(k-1)A - A},\quad\quad
    \cos\theta^{(3)}_{3k} = \frac{(-1)^{k-1}2A}{2+3(k-1)A + A}. 
\end{align}
The corresponding OPUC are 
\begin{subequations}
\label{Eq: P m=3}
\begin{align}
    &P^{(3)}_{3k-2}(z)  = \left(\prod_{j=1}^{3k-2} \frac{1}{\sin\theta^{(3)}_j} \right)\left[z^{3k-2} + (-1)^{k}\cos\theta^{(3)}_{3k-2}\left(\sum_{j=0}^{k-1}z^{3j}\right) + (-1)^{k-1}\frac{\cos\theta^{(3)}_{3k-3}\cos\theta^{(3)}_{3k-2}}{\cos\theta^{(3)}_{3k}} \left(\sum_{j=1}^{k-1}z^{3j-1}\right)  \right.\nonumber\\
    &\phantom{p_{3k-2} =} \left. +(-1)^k \frac{\cos\theta^{(3)}_{3k-3}}{2\cos\theta^{(3)}_{3k-1}}\left( \cos\theta^{(3)}_{3k-2} - \cos\theta^{(3)}_{3k-1} \right)\left( \sum_{j=1}^{k-1}z^{3j-2} \right) \right];\\
    &P^{(3)}_{3k-1}(z) = \left(\prod_{j=1}^{3k-1} \frac{1}{\sin\theta^{(3)}_j} \right)\left[z^{3k-1} + (-1)^{k-1}\cos\theta^{(3)}_{3k-1}\left( \sum_{j=0}^{k-1}z^{3j} + z^{3j+1} \right) - 2(-1)^{k-1}\cos\theta^{(3)}_{3k-1}\left( \sum_{j=1}^{k-1} z^{3j-1} \right) \right];\\
    &P^{(3)}_{3k}(z) = \left(\prod_{j=1}^{3k} \frac{1}{\sin\theta^{(3)}_j} \right)\left[z^{3k} + (-1)^{k-1}\frac{\cos\theta^{(3)}_{3k}}{2}\left( \sum_{j=0}^{k-1}z^{3j+1} + z^{3j+2} \right) -(-1)^{k-1}\cos\theta^{(3)}_{3k}\left( \sum_{j=0}^{k-1}z^{3j} \right) \right].
\end{align}
\end{subequations}

\end{widetext}
The corresponding spectral function is
\begin{align}
    \Phi^{(3)}(e^{i\omega}) = \frac{1}{2}\left[ \Phi^{(1)}(e^{i(\omega-2\pi/3)}) +  \Phi^{(1)}(e^{i(\omega+2\pi/3)})\right].
\end{align}

For $m = 6$, $G_L^{(6)}(z)$ is 
\begin{align}
    G_L^{(6)}(z) = \frac{1-A - \left(1-\frac{A}{2}\right)z+z^2}{1-z+z^2}.
\end{align}
Similar to the relation between the $m = 1,2$ cases, we have $G_L^{(6)}(z) = G_L^{(3)}(-z)$. Therefore, the Krylov angles, OPUC and spectral function satisfy
\begin{align}
    &\theta^{(6)}_k = \bigg\{\begin{array}{cc}
    \pi - \theta^{(3)}_k & \text{if}\ k\ \text{is odd};\\
     \theta^{(3)}_k & \text{else}.
    \end{array} \label{Eq: m=6 Krylov angle}\\
    &P^{(6)}_k(z) = (-1)^k P^{(3)}_k(-z); \label{Eq: P m=6}\\
    &\Phi^{(6)}(e^{i\omega}) = \frac{1}{2}\left[ \Phi^{(1)}(e^{i(\omega-\pi/3)}) +  \Phi^{(1)}(e^{i(\omega+\pi/3)})\right].
\end{align}

In Fig.~\ref{Fig: m-period OPUC}, we present numerical results for $1/|P^{(m)}_k(e^{i\omega})|^2$ for $m = 1,2,3,4,6$ and show consistency with the spectral function in the large $k$ limit according to \eqref{Eq: Bernstein large k}.

\section{Numerical results for the $Z_3$ clock model and the 6-sublattice chain} \label{AppZ3}
In this section, we first summarize the results for the $Z_3$ clock model with open boundary conditions appearing in \cite{sreejith2016parafermion}. The model was studied for two different parameters: one corresponded to long-lived edge mode resembling the $3$-period autocorrelation \eqref{Eq: A m-period}, and the other corresponded to an edge mode that decayed at late times. We derive an exactly solvable 
Floquet ITFIM that captures this behavior.  In particular, we show that a 6-period spatial pattern in Krylov angles leads to 3-period dynamics. 

\begin{figure*}
    \includegraphics[width=0.31\textwidth]{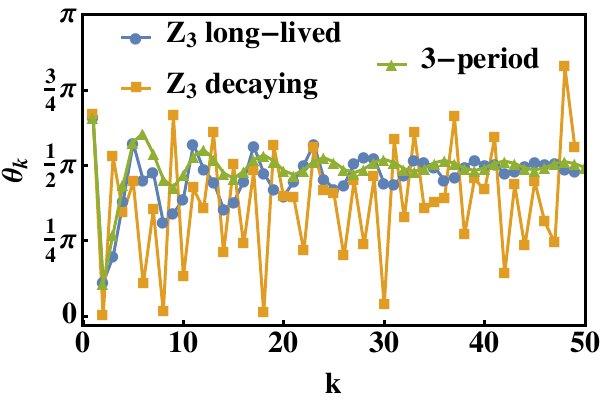}
    \includegraphics[width=0.32\textwidth]{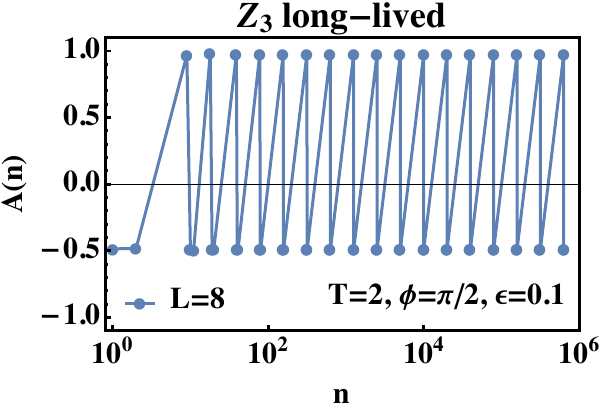}
    \includegraphics[width=0.32\textwidth]{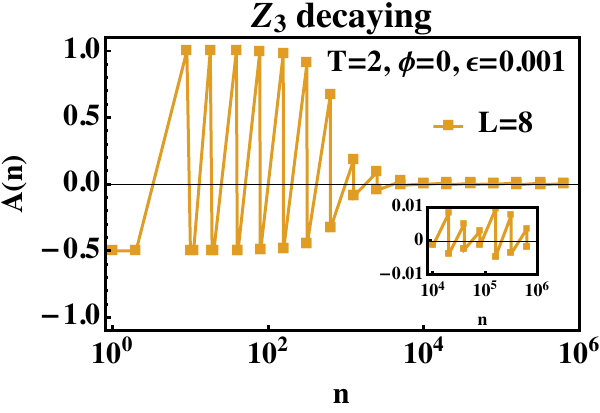}
    \includegraphics[width=0.4\textwidth]{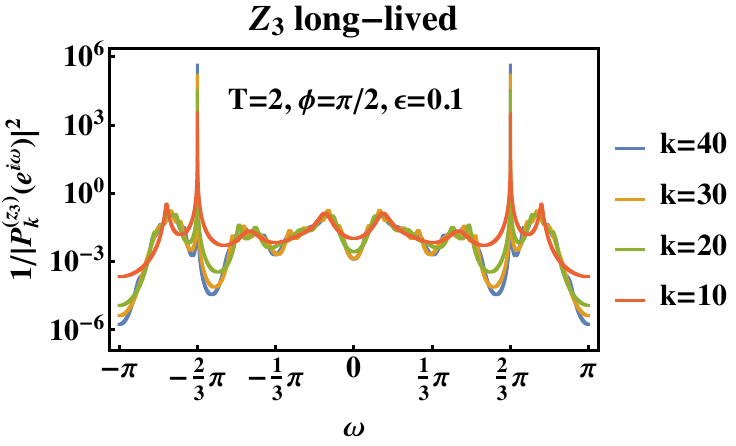}
    \includegraphics[width=0.4\textwidth]{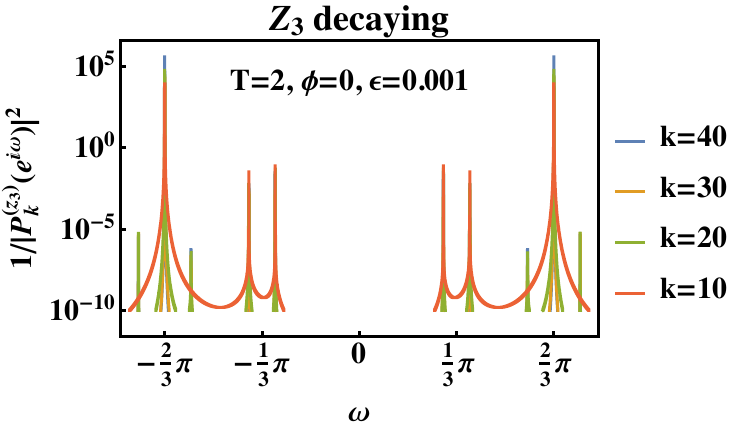}    

    \caption{Numerical results of for the Krylov angles, autocorrelation and OPUC for $Z_3$ clock model \eqref{Eq: Z3 clock model} with initial operator   $(\sigma_1 + \sigma_1^\dagger)/\sqrt{2}$ and couplings $T=2$, $J=e^{i\phi}$, $g = \epsilon + i2\pi/(3\sqrt{3})$. First row: the left panel shows Krylov angles for $L = 8$ for long-lived ($T=2, \phi = \pi/2, \epsilon=0.1$) and decaying ($T=2, \phi = 0, \epsilon=0.001$) cases. The Krylov angles of the 3-period case \eqref{Eq: m=3 Krylov angle} are also plotted for comparison and we set its first Krylov angle to match the $Z_3$ long-lived result. The Krylov angles for both the  long-lived and 3-period cases show convergence to $\pi/2$ for large $k$ along with a 6-periodic spatial oscillation in $k$. However, no simple pattern is observed for the decaying case. The corresponding autocorrelation of $Z_3$ clock model are shown in the middle and right panels separately. The upper (lower) envelope in the upper middle and upper right panels correspond to $n+2\,  (n,n+1)$ where $n$ equally slices the time in the logarithmic-scale and $n = 1\ \text{mod}\ 3$. Interestingly, the decaying autocorrelation shows a different periodicity at late-times (see inset). Second row: The numerical results for $1/|P_k(e^{i\omega})|^2$ for long-lived (left panel) and decaying (right panel) cases for $k = 10, 20, 30, 40$. The peaks at $\pm 2\pi/3$ reflect the 3-period oscillation of the autocorrelation in both cases. There are peaks at other frequencies for the decaying case as indicated by the late-time oscillation of the autocorrelation. Numerically, it is subtle to conclude if the peaks will eventually lead to delta functions in the spectral function, see discussion in the main text.}
    \label{Fig: Z3}
\end{figure*}

\subsection{$Z_3$ clock model}
The $Z_3$ clock model with open boundary conditions \cite{sreejith2016parafermion} is given by the following unitary evolution 
\begin{align}
    U_{z_3} = e^{-i\frac{T}{2}H_g}e^{-i\frac{T}{2}H_J},
    \label{Eq: Z3 clock model}
\end{align}
where
\begin{align}
    &H_J = J\sum_{i=1}^{ L-1} \sigma_i\sigma_{i+1}^\dagger + \text{H.c.}; &H_g = g\sum_{i=1}^L \tau_i + \text{H.c.},\\
    &\sigma = \begin{pmatrix}
        1 & 0 & 0\\
        0 & e^{i2\pi/3} & 0\\
        0 & 0 & e^{i4\pi/3}
    \end{pmatrix};
    &\tau = \begin{pmatrix}
        0 & 0 & 1\\
        1 & 0 & 0\\
        0 & 1 & 0
    \end{pmatrix}.
\end{align}
We consider the autocorrelation of the edge hermitian operator $(\sigma_1 + \sigma_1^\dagger)/\sqrt{2}$ with the following parametrization $T=2$, $J=e^{i\phi}$, $g = \epsilon + i2\pi/(3\sqrt{3})$. The $Z_3$ clock model has an edge mode completely localized on the first site for $\epsilon=0$. As one moves away from $\epsilon=0$, the existence of the edge mode depends on the phase $\phi$ of $J$   \cite{sreejith2016parafermion} but so far an analytic expression is known only in the high frequency limit \cite{Fendley2012,jermyn2014stability}. Two setups are presented in Fig.~\ref{Fig: Z3}: (i) long-lived, period tripled edge mode  at $\phi=\pi/2$, $\epsilon =0.1$. (ii) 
short-lived, period tripled edge mode at $\phi=0$, $\epsilon = 0.001$. 

The Krylov angles for these two setups are shown in the left panel and the corresponding autocorrelations are shown in the middle and right panels of the first row of Fig.~\ref{Fig: Z3}. The orthogonal Krylov basis is computed by the Gram-Schmidt procedure numerically, with the angles calculated from the matrix elements \eqref{Eq: upper-Hessenberg} and the relations \eqref{Eq: Verblunsky Krylov angle}. The Krylov angles of the 3-period case \eqref{Eq: m=3 Krylov angle} are plotted as well for comparison and the $A$ is chosen to coincide with the first Krylov angle of the long-lived edge mode of the  $Z_3$ model. 

The two setups have completely different behavior of the Krylov angles: the angles approach $\pi/2$ for the the long-lived case but fluctuate considerably for the decaying case. Note that this observation of fluctuation in the decaying case is special for the $Z_3$ clock model since the Krylov angles in $Z_2$  models typically approach $\pi/2$ for decaying autocorrelations, see discussion in Ref.~\cite{yeh2023universal,yeh2025moment}. Although the Krylov angles for the $Z_3$ long-lived and 3-period case are not identical, they share a similar oscillation pattern of period $6$ in the spatial index.  

The Bernstein–Szeg\H{o} approximation results are presented in the second row of Fig.~\ref{Fig: Z3}, where the OPUC are calculated from the Szeg\H{o} recurrence relation \eqref{Eq: Szego} with numerically determined Krylov angles. Both plots show peaks at $\pm 2\pi/3$ as the autocorrelations have a periodicity of 3 for sufficiently long times. Interestingly, the $Z_3$ decaying result also has some additional peaks reflecting the late time oscillations shown in the inset in the lower right panel in Fig.~\ref{Fig: Z3}. However, it is subtle to conclude that the peaks will lead to delta functions numerically. 

This point can be elaborated by writing the OPUC as follows
\begin{align}
    P_k(z) = \kappa_{kk}p_k(z);\ \kappa_{kk} = \prod_{j=1}^k\frac{1}{\sin\theta_j},
    \label{Eq: OPUC decomposition}
\end{align}
where $p_k(z)$ is the monic polynomial whose prefactor of $z^k$ is fixed to be 1.  According to the Bernstein–Szeg\H{o} approximation \eqref{Eq: Bernstein approx} and \eqref{Eq: Bernstein large k}, one would expect that in order to have a delta function peak in the spectral function, $P_k(z)$ should approach zero at some frequency for large $k$ so that its inverse diverges. From the decomposition \eqref{Eq: OPUC decomposition}, whether $P_k(z)$ approaches zero depends on the competition between $\kappa_{kk}$ and $p_k(z)$. For the cases where the Krylov angles approach $\pi/2$ such as for the $m$-period case in Fig.~\ref{Fig: m-period OPUC} and the $Z_3$ long-lived case in Fig.~\ref{Fig: Z3}, the numerics are more stable as $\kappa_{kk}$ is constant for large enough $k$, with the peaks of the approximate spectral function growing with $k$. On the other hand, if the Krylov angles do not converge to $\pi/2$, $\kappa_{kk}$ will diverge for large $k$. Therefore, the peak in the spectral function can only arise  from cancellations between small $p_k(z)$ and large $\kappa_{kk}$ which is a numerically unstable scenario. For example, numerical results of $\kappa_{kk}$ for decaying $Z_3$ autocorrelator are about $8\times 10^4, 8\times 10^6, 3\times 10^8, 9\times 10^8$ for $k=10,20,30,40$ and the  maximum value of the peaks for the spectral function for this case do not grow with $k$, see Fig.~\ref{Fig: Z3}. Therefore, one requires further numerical or analytic studies to resolve the delta functions, if any, for this case. 

Nevertheless, the peaks in the approximate spectral functions do provide hints for possible long-lived modes. Moreover, we should stress that $P_k(z)$ is constructed from the dynamics in the first $k$ time steps. 

An important physical consequence of a divergent $\kappa_{kk}$ is a gap in the spectral function. In Fig.~\ref{Fig: Z3}, except for some handful of frequencies, the approximate spectral function for decaying $Z_3$ autocorrelator becomes negligible (less than $10^{-10}$) at most frequencies. This indicates a gap in the spectral function in the large $k$ limit.

We conclude this section by making an interesting observation that the autocorrelation for long-lived and decaying cases are similar for $n < 10^2$ but the Krylov angles and the orthogonal polynomials already show huge differences for small $k$ in Fig.~\ref{Fig: Z3}. 

\subsection{6-sublattice Majorana chain}
So far, we have numerically examined the three-period dynamics in the $Z_3$ clock model. Since the analytic expression for the edge modes in this model is only available in the high-frequency limit, and that to for some special couplings \cite{Fendley2012}, we instead study the ITFIM to reconstruct the 3-period dynamics and derive its analytic solutions. 

Note that the physical model is a $Z_3$ spin chain, while the ITFIM mapping is to a $Z_2$ chain. The latter has zero and $\pi$ long lived modes occurring naturally. However to accommodate the long lived modes of a $Z_3$ chain with a time-periodicity of $2\pi/3$, one needs fine-tuned Krylov angles. Here we show that this fine tuning involves a spatial periodicity of $6$ in the Krylov angles.

\begin{figure*}
    \includegraphics[width=0.23\textwidth]{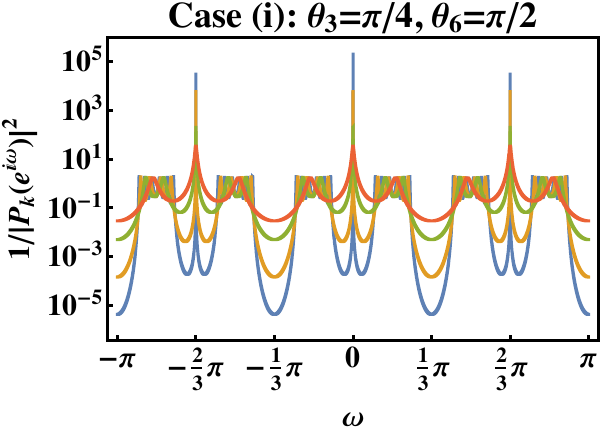}
    \includegraphics[width=0.23\textwidth]{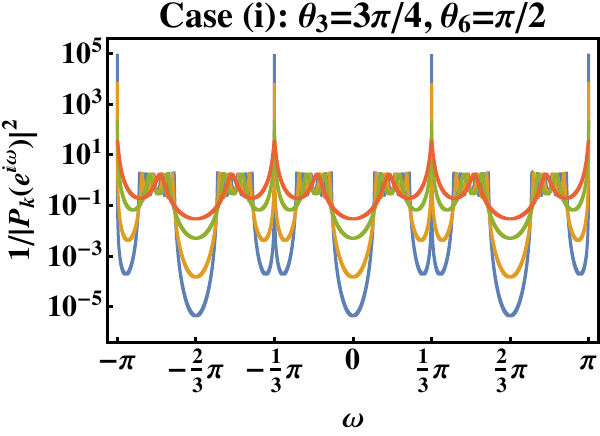}
    \includegraphics[width=0.23\textwidth]{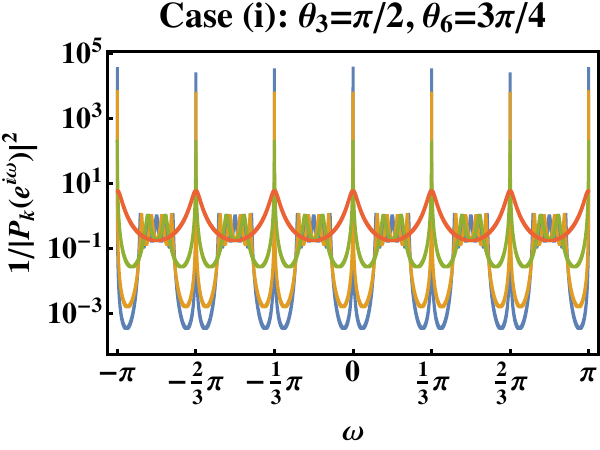}
    \includegraphics[width=0.28\textwidth]{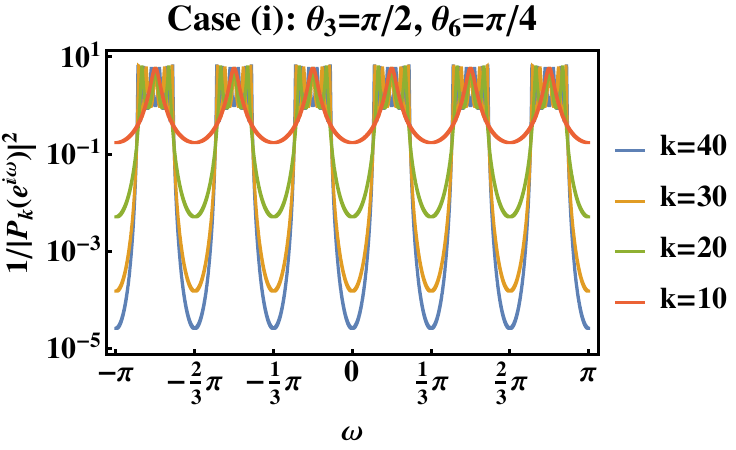}
    \includegraphics[width=0.23\textwidth]{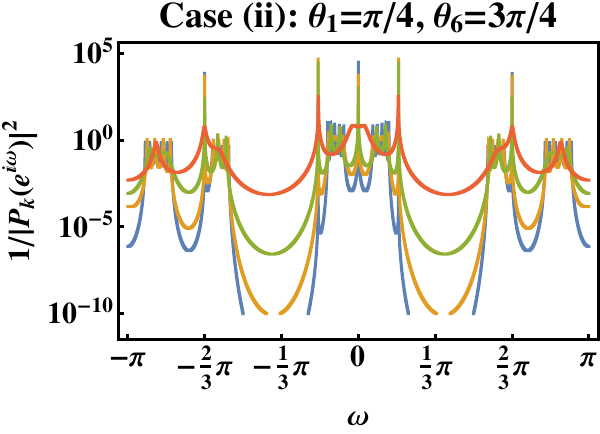}
    \includegraphics[width=0.23\textwidth]{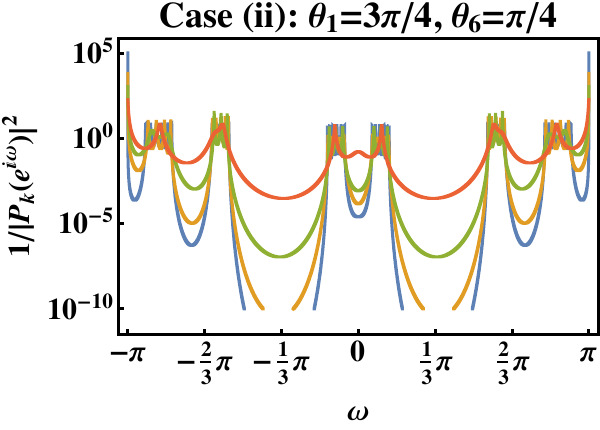}
    \includegraphics[width=0.23\textwidth]{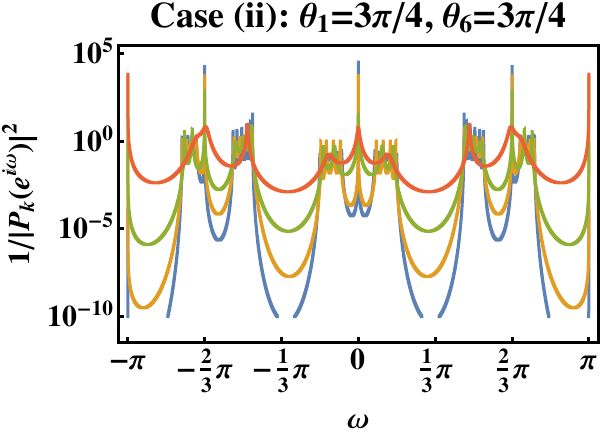}
    \includegraphics[width=0.28\textwidth]{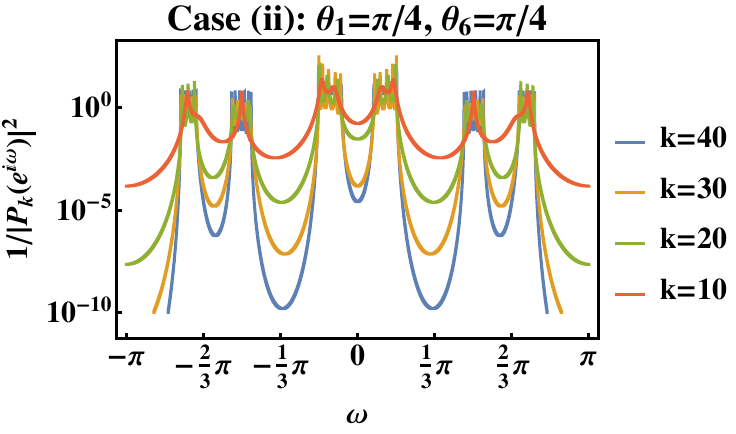}
    \includegraphics[width=0.23\textwidth]{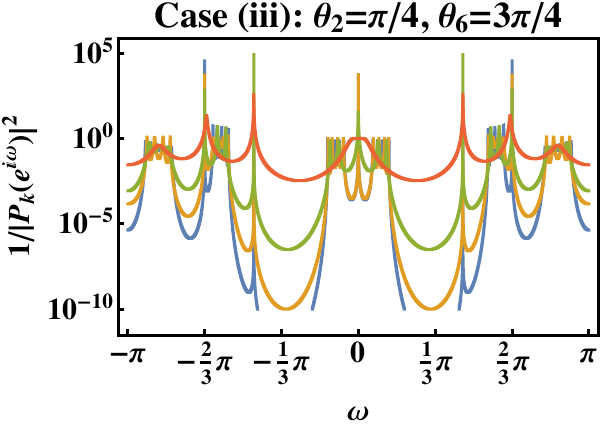}
    \includegraphics[width=0.23\textwidth]{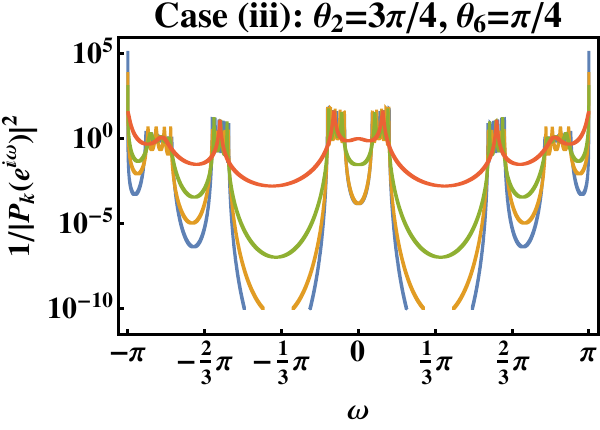}
    \includegraphics[width=0.23\textwidth]{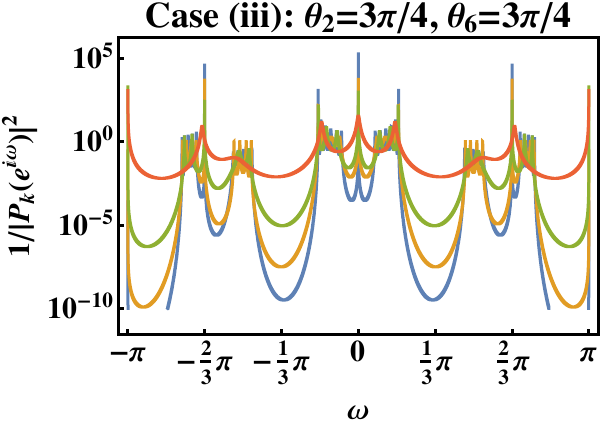}
    \includegraphics[width=0.28\textwidth]{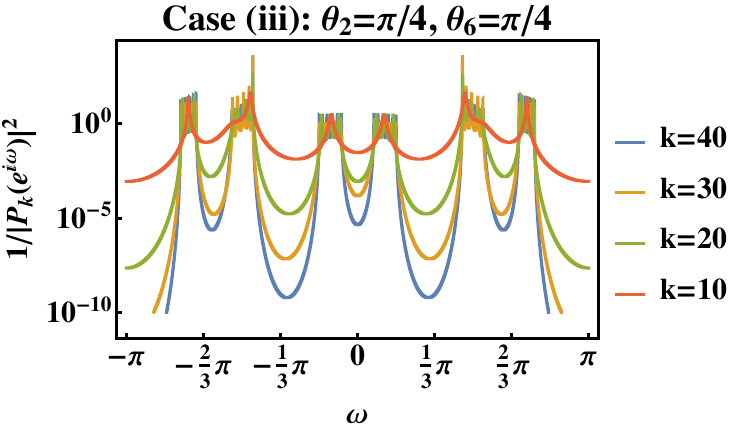} 

    \caption{The numerical results of $1/|P_k(e^{i\omega})|^2$ for the 6-sublattice Majorana chain with the Krylov angles \eqref{Eq: 6-sublattice angles}. The approximate spectral function captures the dynamics of $\gamma_1$ with its peaks corresponding to long-lived modes that overlap with $\gamma_1$ The first row shows results for the case (i): $\theta_1 = \theta_2 = \theta_4 = \theta_5 = \pi/2$ with $\theta_3, \theta_6$ being free parameters. From left to right panels, we set different values of $\theta_3, \theta_6$ such that it shows four different scenarios: $0,\ \pm 2\pi/3$-modes; $\pi,\ \pm\pi/3$-modes; all six modes; no modes. The $0$ and $\pm 2\pi/3$-modes ($\pi$ and $\pm \pi/3$-mode) always appear together because of the threefold degeneracy of the eigenvalue spectrum \eqref{Eq: K R commutation}. The second and third rows respectively show results for cases (ii): $\theta_2 = \theta_3 = \theta_5 = \pi/2$ with $\theta_1=\theta_4, \theta_6$ being free parameters, and (iii): $\theta_1 = \theta_3 = \theta_4 = \pi/2$ and $\theta_2=\theta_5, \theta_6$ being free parameters. In contrast to case (i), cases (ii) and (iii) do not have a threefold degenerate spectrum (see Fig.~\ref{Fig: Folded spectrum}), which leads to the absence of the $\pm \pi/3$-mode. The Krylov angles for cases (ii) and (iii) from left to right panels in second and third rows are evaluated for four different situations: $0,\ \pm 2\pi/3$-modes; only $\pi$-modes; $0,\ \pi, \pm2\pi/3$-modes; absence of $0, \pm2\pi/3, \pi$-modes. Note that there are other modes for cases (ii) and (iii) as indicated by the peaks not located at $\omega = 0,\pm2\pi/3, \pi$ in the second and third rows. These new modes can be viewed as deformation of $\pm\pi/3$-mode from case (i) (see Fig.~\ref{Fig: Deformation spectrum}).}
    \label{Fig: 6-sublattice OPUC}
\end{figure*}

This can be explicitly seen in Fig.~\ref{Fig: Z3} where the Krylov angles of the autocorrelations of the $Z_3$ clock model show a 6-site periodic structure,  along with the expected $2\pi/3$ time-periodicity. Based on this observation, we now study the ITFIM \eqref{Eq: ITFIM} with perfect 6-periodic Krylov angles
\begin{align}
    \theta_{6k+j} = \theta_j,\ \forall 1 \leq j \leq 6,\ k\geq 0.
    \label{Eq: 6-sublattice angles}
\end{align}
From the Jordan-Wigner transformation \eqref{Eq: Jordan-Wigner}, this model becomes a Majorana chain with a 6-sublattice structure. To solve for the Krylov angles that support a $2\pi/3$-mode, we consider the eigen-operator equation: $K_I|\Psi_{2\pi/3}) = e^{i2\pi/3}|\Psi_{2\pi/3})$, where $\Psi_{2\pi/3}$ is a linear combination of Majoranas $\gamma_j$, $\Psi = \sum_{j}\psi_j\gamma_j$ and $K_I$ in the Majorana basis is given by \eqref{Eq: Majorana basis matrix element}. In contrast to $0$ and $\pi$-modes, the coefficients $\psi_j$ for $\Psi_{2\pi/3}$ can be complex. Therefore, $\Psi_{2\pi/3}$ is normalized according to $\Psi_{2\pi/3}\Psi_{2\pi/3}^\dagger = \mathbb{I}$. Note that the $(-2\pi/3)-$mode can be obtained from Hermitian conjugation, $\Psi_{-2\pi/3} = \Psi_{2\pi/3}^\dagger$, and therefore one only needs to solve $\Psi_{2\pi/3}$.  Explicitly, the eigen-operator equation leads to (see Appendix \ref{App: J})
\begin{widetext}
\begin{subequations}
\label{Eq: eigen-operator equ}
\begin{align}
    &\psi_2 = \frac{e^{i\frac{2\pi}{3}}-\cos\theta_1}{\sin\theta_1} \psi_{1};\\
    &\psi_j = \frac{e^{-i\frac{(-1)^j2\pi}{3}}\sin\theta_{j-2}}{\sin\theta_{j-1}}\psi_{j-2} + (-1)^j\frac{e^{-i\frac{(-1)^j2\pi}{3}}\cos\theta_{j-2}-\cos\theta_{j-1}}{\sin\theta_{j-1}} \psi_{j-1},\ \forall j \geq 3.
\end{align}
\end{subequations}
\end{widetext}
 Since we assumed the 6-sublattice structure of the Majorana chain, we make the following conjecture about the coefficient of the eigen-operator $\Psi_{2\pi/3}$ 
\begin{align}
    \psi_{6k+j} = c_j \xi^k,\ \forall \, 1 \leq j \leq 6,\ k\geq 0.
    \label{Eq: 6-sublattice psi}
\end{align}
We solve \eqref{Eq: 6-sublattice angles}, \eqref{Eq: eigen-operator equ} and \eqref{Eq: 6-sublattice psi} self-consistently and numerically (see Appendix \ref{App: J}). We find the following three types of solutions: (i) $\theta_1 = \theta_2 = \theta_4 = \theta_5 = \pi/2$ and $\xi = \tan(\theta_3/2)\cot(\theta_6/2)$, (ii) $\theta_2 = \theta_3 = \theta_5 = \pi/2$, $\theta_1=\theta_4$ and $\xi = \cot(\theta_6/2)$, (iii) $\theta_1 = \theta_3 = \theta_4 = \pi/2$, $\theta_2=\theta_5$ and $\xi = \cot(\theta_6/2)$. The analytic solutions for the $2\pi/3$-mode \eqref{Eq: eigen-operator equ} are presented in Appendix \ref{App: J}.

\begin{figure*}
    \includegraphics[width=0.32\textwidth]{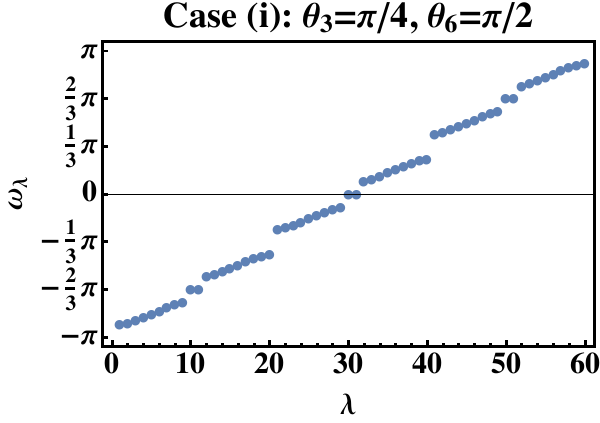}
    \includegraphics[width=0.32\textwidth]{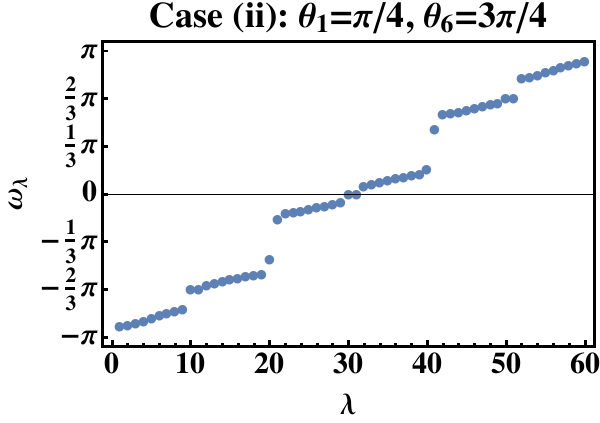}
    \includegraphics[width=0.32\textwidth]{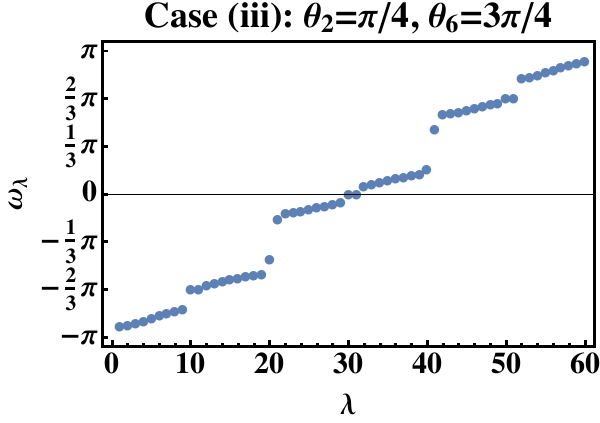}
    \includegraphics[width=0.32\textwidth]{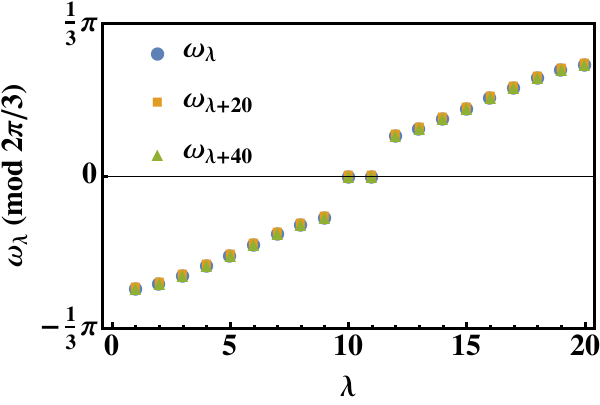}
    \includegraphics[width=0.32\textwidth]{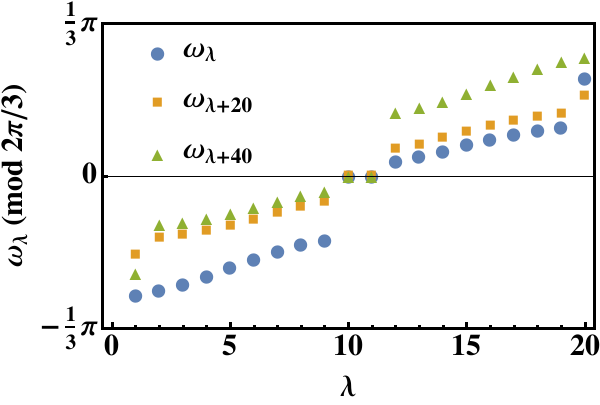}
    \includegraphics[width=0.32\textwidth]{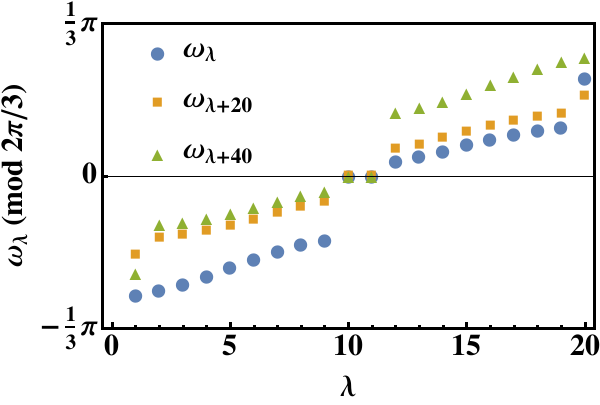} 

    \caption{The single particle eigenvalue spectrum from ED of the 6-sublattice Majorana chain with the Krylov angles \eqref{Eq: 6-sublattice angles} and for a system size of 60-Majoranas. We choose the parameters to be the same as the leftmost three columns of Fig.~\ref{Fig: 6-sublattice OPUC}. The full ED spectra is presented in the first rows and the corresponding folded spectra are in the corresponding lower panels. The threefold degeneracy of case (i) agrees with the algebraic relation \eqref{Eq: K R commutation}. In contrast, cases (ii) and (iii) do not have such a property. Therefore, it leads to the missing $\pm \pi/3$-modes in the second and third rows of Fig.~\ref{Fig: 6-sublattice OPUC}. Note that the spectra in the middle and right panels are exactly the same because cases (ii) and (iii) are related by reflection of the Majorana chain.}
    \label{Fig: Folded spectrum}
\end{figure*}

\begin{figure}
    \includegraphics[width=0.32\textwidth]{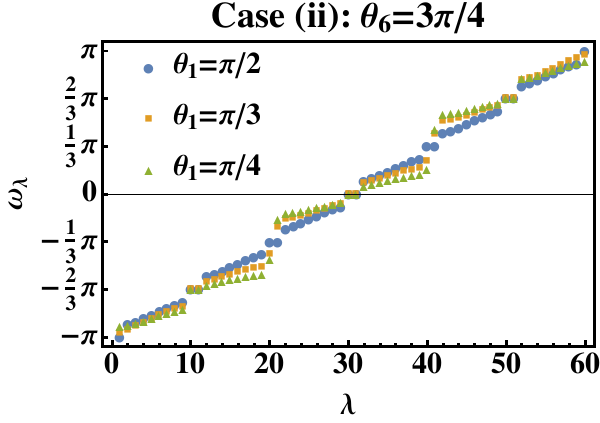}

    \caption{The eigenvalue spectrum from ED of the 6-sublattice Majorana chain with a system size of 60-Majoranas and for case (ii): $\theta_2 = \theta_3 = \theta_5 = \pi/2$, $\theta_1=\theta_4$. We consider $\theta_1 = \pi/2, \pi/3, \pi/4$ separately and fixed $\theta_6 = 3\pi/4$. Changing $\theta_1$ from $\pi/2$ to $\pi/4$ can be viewed as a deformation from the third panel of case (i) to the leftmost panel of case (ii) in Fig.~\ref{Fig: 6-sublattice OPUC}. At $\theta_1 = \pi/2$, $\pm \pi/3$-modes exist. As $\theta_1$ decreases, the eigenvalues at $\pm \pi/3$ split and turn into new modes at different frequencies.}
    \label{Fig: Deformation spectrum}
\end{figure}

The spectral function captures the dynamics of $\gamma_1$ and therefore reveals all possible long-lived modes that have overlap with $\gamma_1$.
For case (i), Fig.~\ref{Fig: 6-sublattice OPUC} shows the numerical results for $1/|P_k(e^{i\omega})|^2$ in the first row for the following four different scenarios from left to right: 0 and $\pm 2\pi/3$-modes; $\pi$ and $\pm\pi/3$-modes; all six modes; and no modes. Note that the $\pm 2\pi/3$-modes ($\pm \pi/3$-mode) coexist with the $0$-mode ($\pi$-mode) due to the following property of the superoperator $K_I$ 
\begin{align}
    K_IR = e^{i\frac{2\pi}{3}}RK_I;\ K_IR^{-1} = e^{-i\frac{2\pi}{3}}R^{-1}K_I
    \label{Eq: K R commutation}
\end{align}
where $R$ has the following diagonal form in the Majorana basis 
\begin{align}
    &(\gamma_j|R|\gamma_k)  = \begin{pmatrix}
        r &\\
        & r\\
        & & \ddots
    \end{pmatrix};\\
    &r = \begin{pmatrix}
        1 &\\
        & e^{i\frac{2\pi}{3}}\\
        & & e^{-i\frac{2\pi}{3}}\\
        & & & e^{-i\frac{2\pi}{3}}\\
        & & & & e^{i\frac{2\pi}{3}}\\
        & & & & & 1
    \end{pmatrix}.
\end{align}
 The relation \eqref{Eq: K R commutation} implies that any eigen-operator $|\Psi_\lambda)$ of $K_I$ with eigenvalue $e^{i\omega_\lambda}$ must be associated  with the eigen-operators $R|\Psi_\lambda),\ R^{-1}|\Psi_\lambda)$  with eigenvalues $e^{i(\omega_\lambda + 2\pi/3)},\ e^{i(\omega_\lambda - 2\pi/3)}$ respectively.  Therefore, if one considers $\omega_\lambda$ mod $2\pi/3$, the folded spectrum is three-fold degenerate, see left column in Fig.~\ref{Fig: Folded spectrum}. 
 We perform exact diagonalization (ED) for $\theta_3 = \pi/4,\ \theta_6= \pi/2$ (same setting as left panel in the first row of Fig.~\ref{Fig: 6-sublattice OPUC}) for a system size of 60-Majoranas. The lower panel demonstrates the three-fold degeneracy of $\omega_\lambda$ mod $2\pi/3$ agreeing with the existence of the rotation matrix $R$. Note that $K_I$ \eqref{Eq: K matrix}  describes Floquet dynamics in the single Majorana basis and therefore its eigenvalues give the single particle spectrum. 
Thus these degeneracies should not be confused with the degeneracies encountered in the many-particle spectra when there are zero modes. 

The cases (ii) and (iii) are related by reflection with respect to the center of the chain. For example, couplings for cases (ii) and (iii) with 12 Majoranas are respectively: $\{ \theta_1, \pi/2, \pi/2, \theta_1, \pi/2, \theta_6, \theta_1, \pi/2, \pi/2, \theta_1, \pi/2 \}$ and $\{\pi/2, \theta_2, \pi/2 , \pi/2, \theta_2, \theta_6, \pi/2, \theta_2, \pi/2, \pi/2, \theta_2\}$. One can easily see that the second set of couplings can be obtained from the first set by reflection and relabeling $\theta_1$ by $\theta_2$. In contrast to case (i), \eqref{Eq: K R commutation} does not hold for cases (ii) and (iii). Therefore, the $\pm \pi/3$ mode is absent in the numerical results presented in the  
second and third rows of Fig.~\ref{Fig: 6-sublattice OPUC}. This difference can be understood as a lack of threefold degeneracy of $\omega_\lambda$ mod $2\pi/3$ for cases (ii) and (iii).  

Fig.~\ref{Fig: Folded spectrum} shows spectra for the same values of Krylov angles as the results shown in the leftmost column of Fig.~\ref{Fig: 6-sublattice OPUC} for the three cases. Note that the missing $\pm \pi/3$-modes in cases (ii) and (iii) are replaced by new modes as seen in the second and third rows of Fig.~\ref{Fig: 6-sublattice OPUC}. In Fig.~\ref{Fig: Deformation spectrum} we numerically show that the new modes are deformations of the $\pm\pi/3$-modes of case (i).  This figure demonstrates the deformation from case (i) with $\theta_1 = \pi/2,\ \theta_6=3\pi/4$ (the third panel of the first row in Fig.~\ref{Fig: 6-sublattice OPUC}) to case (ii) with $\theta_1=\pi/4, \theta_6=3\pi/4$ (the leftmost panel of the second row in Fig.~\ref{Fig: 6-sublattice OPUC}). Note that cases (ii) and (iii) are related by the reflection of the Majorana chain about the center, and therefore the two cases possess the same long-lived modes. The peaks in Fig.~$\ref{Fig: 6-sublattice OPUC}$ only detect the long-lived modes having overlap with the first Majorana. That is the reason why, despite cases (ii) and (iii) being related by reflection symmetry, the peak positions are not identical for the two cases in Fig.~$\ref{Fig: 6-sublattice OPUC}$ as only modes localized on one end of the chain are detected.  

For comparison, we present the results for OPUC for the conventional Floquet transverse-field Ising model in Appendix \ref{App: K}. They show similar features as case (i) in Fig.~\ref{Fig: 6-sublattice OPUC}. This is because the Krylov angles of the two cases are simply related. In particular, the angles in case (i) are equivalent to taking an Ising model with only two angle $\theta_3,\theta_6$, and inserting additional angles $\theta_{1,2,4,5}=\pi/2$.

\section{Conclusions} \label{Conc}

The connection between Krylov complexity and OPUC is a promising route to construct exactly solvable toy models that mimic a range of behavior such as integrable dynamics, chaotic dynamics, and slow decay of quasi-conserved quantities. In this paper, we generalize this mapping to stroboscopic operator dynamics. Such dynamics can be generated by a unitary or by a Hamiltonian where for the latter the dynamics is studied at stroboscopic times.

Remarkably, operator spreading can be mapped to an iconic single particle problem, the ITFIM \cite{yeh2023universal,yeh2025moment}. We connect this result to parallel results obtained using OPUC, where the basis analogous to the ITFIM is known as the CMV basis \cite{CANTERO200329,CANTERO200540,kolganov2025streamlinedkrylovconstructionclassification}. We also establish connections between the Szeg\H{o} approximation for the autocorrelation function and the continued fraction expression for the Laplace transformed autocorrelation function. 

We present several applications. We construct exact OPUC for some simple $m$-periodic dynamics. The last appendix constructs the OPUC for the the Floquet transverse field Ising model in its various topological phases as well as at the dual-unitary point.

We also numerically construct the OPUC for the $Z_3$ clock model. In order to understand the results better, we use the mapping to the ITFIM to construct an exactly solvable model that have the same coarse-grained Krylov parameters as the numerically studied model.

A promising avenue of research is to further develop exactly solvable models using OPUC that have desirable Krylov subspace features inspired by complex quantum systems.

\section*{Acknowledgments}
The authors thank  Dmitrii Trunin for helpful discussion. 
This work was supported by the US Department of Energy, Office of
Science, Basic Energy Sciences, under Award No.~DE-SC0010821.

\section*{Data Availability}
Data and Mathematica codes are available on Zenodo \cite{zenodo}.

\appendix
\section{Derivation of \eqref{Eq: K-matrix in P} and \eqref{Eq: P z^m inner prod}}
\label{App: A}
We first derive relation \eqref{Eq: K-matrix in P}. From \eqref{Eq: linear trans two basis}, one has
\begin{align}
    (\mathcal{O}_k |K| \mathcal{O}_p) &=  \sum_{n=0}^{k}\sum_{m=0}^{p} \overline{\kappa_{kn}} \kappa_{pm} (O(n)|K|O(m)).
\end{align}
By $(O(n)|K|O(m)) = A(m+1-n)$ and applying Fourier transformation \eqref{Eq: Fourier-1}, one obtains
\begin{align}
    &(\mathcal{O}_k |K| \mathcal{O}_p)\nonumber\\
    &= \sum_{n=0}^{k}\sum_{m=0}^{p} \overline{\kappa_{kn}} \kappa_{pm} \int_{-\pi}^{\pi}d\omega \Phi(e^{i\omega}) e^{i\omega (m+1-n)}\nonumber\\
    &= \langle P_k(z),  zP_p(z)\rangle,
\end{align}
where the $z$ in the inner product comes from the extra $e^{i\omega}$ in the Fourier transformation. In the last line, we apply the definition of OPUC \eqref{Eq: OPUC definition} and the inner product notation \eqref{Eq: polynomials inner product}. This is the result shown in \eqref{Eq: K-matrix in P}.

Next, we consider the proof of \eqref{Eq: P z^m inner prod}. Again, with \eqref{Eq: linear trans two basis}, the following relation holds
\begin{align}
    (\mathcal{O}_k|O(m)) = \sum_{n=0}^k\overline{\kappa_{kn}}(O(n)|O(m)).
\end{align}
By identifying $(O(n)|O(m))$ as $A(m-n)$ and Fourier transforming it, we obtain
\begin{align}
    (\mathcal{O}_k|O(m))
    &= \sum_{n=0}\overline{\kappa_{kn}}\int_{-\pi}^\pi d\omega \Phi(e^{i\omega}) e^{i\omega(m-n)}\nonumber\\
    &= \langle P_k(z), z^m \rangle.
\end{align}
Thus, one arrives at the result in \eqref{Eq: P z^m inner prod}.
\section{Derivation of the Szeg\H{o} recurrence relations}
\label{App: B}
In this part, we mainly follow the derivation presented in Ref.~\cite{simon2005orthogonalpart1}. First, let us figure out the orthogonal condition properties of $P_k(z)$ and $P_k^*(z)$. According to the Gram-Schmidt construction,
\begin{align}
    &(\mathcal{O}_k| O(m)) = 0,\ \forall\ 0 \leq m \leq k-1.
\end{align}
By \eqref{Eq: P z^m inner prod}, one concludes
\begin{align}
    \langle P_k(z), z^{m} \rangle = 0,\ \forall\  0 \leq m \leq k-1. \label{Eq: P orthogonal}
\end{align}
For $P_k^*(z)$, we first consider $\langle z^m ,P_k^*(z)\rangle$ and apply the definition of the inner product \eqref{Eq: polynomials inner product} and *-reverse \eqref{Eq: P*},
\begin{align}
    \langle z^m ,P_k^*(z)\rangle = \oint_{|z|=1} \frac{dz}{iz} \Phi(z)  \overline{z^m} z^k\overline{P_k(1/\overline{z})}.
\end{align}
Note that $\overline{z} = z^{-1}$ inside the inner product since $|z| = 1$. The above expression becomes
\begin{align}
    \langle z^m ,P_k^*(z)\rangle &= \oint_{|z|=1} \frac{dz}{iz} \Phi(z)  \overline{P_k(z)}z^{k-m} \nonumber\\
    &= \langle P_k(z), z^{k-m} \rangle.
\end{align}
Therefore, according to \eqref{Eq: P orthogonal},
\begin{align}
    \langle z^m ,P_k^*(z)\rangle = 0,\ \forall\ 1\leq m \leq k. \label{Eq: P^* orthogonal}
\end{align}
As the $zP_k(z)$ term appears in the Szeg\H{o} recurrence relations \eqref{Eq: Szego}, let us consider its orthogonal properties as well
\begin{align}
    \langle zP_k(z), z^m\rangle &= \oint_{|z|=1}\frac{dz}{iz} \Phi(z) \overline{zP_k(z)} z^m \nonumber\\ 
    &= \oint_{|z|=1}\frac{dz}{iz} \Phi(z) \overline{P_k(z)} z^{m-1}\nonumber\\ 
    &= \langle P_k(z), z^{m-1}\rangle.
\end{align}
Therefore,
\begin{align}
    \langle zP_k(z), z^m\rangle = 0,\ \forall\ 1\leq m \leq k. \label{Eq: zP orthogonal}
\end{align}
To derive key relations for OPUC, it is useful to introduce monic polynomials
\begin{align}
    &p_k(z) = \frac{P_k(z)}{\kappa_{kk}} = z^k + \frac{\kappa_{k,k-1}}{\kappa_{kk}}z^{k-1} + \ldots +\frac{\kappa_{k,0}}{\kappa_{kk}};\label{Eq: monic p}\\
    &p_k^*(z) = z^k \overline{p_k(1/\overline{z})}, \label{Eq: p^*}
\end{align}
where $p_k(z)$ is defined such that the prefactor of the highest power term is 1. Note that \eqref{Eq: P orthogonal}, \eqref{Eq: P^* orthogonal} and \eqref{Eq: zP orthogonal} are also valid for their monic counterparts. Below, we first derive the results for the monic polynomials and the results for the OPUC are presented at the end. 

Note that both $zp_k(z)$ and $p_k^*(z)$ satisfy the same orthogonal property \eqref{Eq: P^* orthogonal} and \eqref{Eq: zP orthogonal}. However, the degree of $zp_k(z)$ is one order higher than $p_k^*(z)$. By considering $zp_k(z) - p_{k+1}(z)$, this new polynomial has the same degree as $p_k^*(z)$ and has the same orthogonal properties. Therefore, $zp_k(z) - p_{k+1}(z)$ and $p_k^*(z)$ must only differ by an overall prefactor $\overline{\alpha_k}$
\begin{align}
    zp_k(z) - p_{k+1}(z) = \overline{\alpha_k} p_k^*(z), 
\end{align}
where $\alpha_k$ is called the Verblunsky coefficient. This is the monic version of \eqref{Eq: Szego P-1}. To obtain the monic version of  \eqref{Eq: Szego P-2}, we consider the *-reverse of the above relation. We first take $z \rightarrow 1/\overline{z}$, then apply complex conjugation, and finally multiply by $z^{k+1}$ on both sides. Namely, we have
\begin{align}
    &z^{k+1}\left[\overline{(1/\overline{z})p_k(1/\overline{z})} - \overline{p_{k+1}(1/\overline{z})} \right] = z^{k+1}\overline{\overline{\alpha_k} p_k^*(1/\overline{z})}.
\end{align}
Using \eqref{Eq: p^*}, it can be further simplified to
\begin{align}
    p_k^*(z) - p_{k+1}^*(z) = \alpha_k z p_k(z).
\end{align}
In summary, we derived the monic Szeg\H{o} recurrence relations
\begin{subequations}
\label{Eq: monic Szego}
\begin{align}
    &p_{k+1}(z) = zp_k(z) - \overline{\alpha_k} p_k^*(z), \label{Eq: monic Szego-1}\\
    &p_{k+1}^*(z) = p_k^*(z) - \alpha_k z p_k(z).\label{Eq: monic Szego-2}
\end{align}
\end{subequations}
One can derive the inverse relations by canceling the $p_k(z)$ or $p_k^*(z)$ term in \eqref{Eq: monic Szego}. By canceling $p_k(z)$ one obtains,
\begin{align}
    \alpha_k p_{k+1}(z) + p_{k+1}^*(z) = \left(1 -|\alpha_k|^2\right) p_k^*(z).
\end{align}
By canceling $p_k^*(z)$ on obtains,
\begin{align}
    p_{k+1}(z) + \overline{\alpha_k}p_{k+1}^*(z) = \left(1 -|\alpha_k|^2\right) zp_k(z). 
\end{align}
In summary, the inverse monic Szeg\H{o} recurrence relations are
\begin{subequations}
\begin{align}
    &\rho_k^2 zp_k(z) = p_{k+1}(z) + \overline{\alpha_k}p_{k+1}^*(z);\label{Eq: monic inverse p-1}\\
    &\rho_k^2 p_k^*(z) = \alpha_k p_{k+1}(z) + p_{k+1}^*(z),
\end{align}
\end{subequations}
where we define $\rho_k = \sqrt{1-|\alpha_k|^2}$. Although we did not mention the inverse relations in the main text, they are crucial for the proofs in the later discussions.

From the monic Szeg\H{o} recurrence relations, we can derive relations for the norm of the monic polynomial. From \eqref{Eq: monic Szego-1}
\begin{align}
    &\langle zp_k(z), zp_k(z)  \rangle \nonumber\\
    &= \langle p_{k+1}(z) + \overline{\alpha_k}p_k^*(z), p_{k+1}(z) + \overline{\alpha_k}p_k^*(z) \rangle \nonumber\\
    &= \langle p_{k+1}(z) , p_{k+1}(z)  \rangle + \langle \overline{\alpha_k}p_k^*(z),  \overline{\alpha_k}p_k^*(z) \rangle,
\end{align}
where the cross term $\langle p_{k+1}, p_k^* \rangle$ vanishes since $p_{k+1}$ is orthogonal to any polynomial with degree lower than $k+1$ due to \eqref{Eq: P orthogonal}. By rearranging the above relation,
\begin{align}
    &\langle p_{k+1}(z) , p_{k+1}(z)  \rangle \nonumber\\
    &= \langle zp_k(z), zp_k(z)  \rangle - \langle \overline{\alpha_k}p_k^*(z),  \overline{\alpha_k}p_k^*(z) \rangle\nonumber\\
    &= \left(1 - |\alpha_k|^2 \right)\langle p_k(z), p_k(z)  \rangle\nonumber\\ 
    &= \rho_k^2 \langle p_k(z), p_k(z) \rangle,
\end{align}
where, in the second line, the $z$ prefactor cancels in the first term and $\langle p_k^*(z), p_k^*(z) \rangle = \langle p_k(z), p_k(z) \rangle$ in the second term since $\overline{z} = 1/z$ inside the inner product. Therefore,
\begin{align}
    \langle p_{k+1}(z) , p_{k+1}(z)  \rangle = \prod_{j=0}^k \rho_j^2,
\end{align}
where $\langle p_{0}(z) , p_{0}(z)  \rangle = 1$ according to the definition of the monic polynomial. From 
this argument, $|\alpha_k| \leq 1$ (therefore, $\rho_k \leq 1$) since the norm is positive definite. From this result, we can determine $\kappa_{kk}$ in \eqref{Eq: monic p}, 
\begin{align}
    \langle  p_k(z),  p_k(z) \rangle = \left\langle \frac{P_k(z)}{\kappa_{kk}}, \frac{P_k(z)}{\kappa_{kk}}\right\rangle = \frac{1}{\kappa_{kk}^2},
\end{align}
where $\langle P_k(z), P_k \rangle = 1$ by definition. Note that the OPUC are defined up to some arbitrary overall complex phase and one can choose $\kappa_{kk}$ to be real and positive. This choice is consistent with the initial condition $P_0(z) = P_0^*(z) = 1$. Therefore, we obtain the following relation for $\kappa_{kk}$
\begin{align}
    \kappa_{kk} = \prod_{j=0}^{k-1} \frac{1}{\rho_j}.\label{Eq: kappa rho}
\end{align}
Combining \eqref{Eq: monic p}, \eqref{Eq: monic Szego} and \eqref{Eq: kappa rho}, the Szeg\H{o} recurrence relations for the OPUC are
\begin{subequations}
\begin{align}
    &\rho_kP_{k+1}(z) = zP_k(z) - \overline{\alpha_k}P_k^*(z);\\
    &\rho_k P_{k+1}^*(z) = P_k^*(z) - \alpha_k z P_k(z).
\end{align}
\end{subequations}
They are exactly the results shown in \eqref{Eq: Szego}. For the inverse relations, we have 
\begin{subequations}
\label{Eq: inverse Szego}
\begin{align}
    &\rho_k zP_k(z) = P_{k+1}(z) + \overline{\alpha_k} P_{k+1}^*(z);\label{Eq: inverse Szego P-1}\\
    &\rho_k P_k^*(z) = \alpha_k P_{k+1}(z) + P_{k+1}^*(z).\label{Eq: inverse Szego P-2}
\end{align}
\end{subequations}

\section{Inner product between OPUC and its $*$-reverse}
\label{App: C}
In this appendix, we will prove the inner product relation between $P_k(z)$ and $P_k^*(z)$ and follow Ref.~\cite{simon2005orthogonalpart1}.  Let us first consider $\langle P_j^*(z), P_k(z) \rangle$. For $k \geq j+1$,
\begin{align}
    \langle P_j^*(z), P_k(z) \rangle = 0,\ \forall\,\,  k \geq j+1
\end{align}
since $P_j^*(z)$ is a polynomial of degree $j$ and $P_k(z)$ is orthogonal to $\{ 1, z^1, z^2, \ldots, z^{k-1} \}$, see \eqref{Eq: P orthogonal}. For $k \leq j$, we have the following relation
\begin{align}
    \langle P_j^*(z), P_k(z) \rangle = \langle P_j^*(z), 1 \rangle P_k(0),\ \forall\,\,  0 \leq k \leq j
    \label{Eq: P^* 1}
\end{align}
Note that $P_j^*(z)$ is orthogonal to $\{z, z^2,\ldots, z^{j} \}$, see \eqref{Eq: P^* orthogonal}. Namely, for any polynomial with degree lower than $j+1$, only the $z^0$ term contributes to the inner product with $P_j^*(z)$. Moreover, we can replace the inner product  $\langle P_j^*(z), 1 \rangle$ by $\langle P_j^*(z), f(z) \rangle$ for arbitrary polynomial $f(z)$ with degree lower than $j+1$ and $f(0) = 1$. Therefore, we can replace the $1$ inside \eqref{Eq: P^* 1} by $P^*_j(z)/\kappa_{jj}$ and then apply $\langle P^*_j(z), P^*_j(z) \rangle = \langle P_j(z), P_j(z) \rangle = 1$. Eventually, we get
\begin{align}
    \langle P_j^*(z), P_k(z) \rangle = \frac{P_k(0)}{\kappa_{jj}},\  \forall\,\, 0 \leq k \leq j.
\end{align}
In order to derive the expression for $P_k(0)$, we evaluate the monic Szeg\H{o} recurrence relations \eqref{Eq: monic Szego-1} at $z=0$,
\begin{align}
    &p_{k+1}(0) = 0\times p_k(0) - \overline{\alpha_k}p_k^*(0) = -\overline{\alpha_k}, 
\end{align}
where $p_k^*(0) = 1$ according to the definition of the monic polynomial \eqref{Eq: p^*}. Therefore,
\begin{align}
    &\langle P_j^*(z), P_k(z) \rangle \nonumber\\ 
    &= \frac{\kappa_{kk}p_k(0)}{\kappa_{jj}} = -\overline{\alpha_{k-1}} \prod_{l=k}^{j-1}\rho_l,\ \text{for } 0\leq k \leq j,
\end{align}
where we apply $p_k(0) = -\overline{\alpha_{k-1}}$ and \eqref{Eq: kappa rho}. In summary,
\begin{align}
    \langle P_j^*(z), P_k(z) \rangle = \left\{\begin{array}{cc}
     -\overline{\alpha_{k-1}} \prod_{l=k}^{j-1}\rho_l, & \text{for } 0\leq k \leq j ;\\
     0, & \text{else}.
     \end{array}\right.
\end{align}
By complex conjugation and exchanging the indices $j,k$, one obtains
\begin{align}
    \langle P_j(z), P_k^*(z) \rangle &= \overline{\langle P_k^*(z), P_j(z) \rangle}\nonumber\\
    &=\left\{\begin{array}{cc}
     -\alpha_{j-1} \prod_{l=j}^{k-1}\rho_l, & \text{for } 0\leq j \leq k ;\\
     0, & \text{else}.
     \end{array}\right.
     \label{Eq: P P^* inner product}
\end{align}
Next, we consider $\langle P_j^*(z), P_k^*(z) \rangle$. Similar to the above discussion, we have the following relation for $0 \leq k \leq j$,
\begin{align}
    \langle P_j^*(z), P_k^*(z) \rangle = \langle P_j^*(z), P_j^*(z) \rangle \frac{\kappa_{kk}p_k^*(0)}{\kappa_{jj}} = \prod_{l=k}^{j-1}\rho_l. 
\end{align}
For $k \geq j+1$, one only needs to exchange the indices $j, k$ in the previous expression
\begin{align}
    \langle P_j^*(z), P_k^*(z) \rangle = \prod_{l=j}^{k-1}\rho_l.
\end{align}
In summary, 
\begin{align}
    \langle P_j^*(z), P_k^*(z) \rangle =  \prod_{l=\text{min}(j,k)}^{\text{max}(j,k)-1}\rho_l.
    \label{Eq: P* P* inner product}
\end{align}
Now, we have all the key relations for the proof of the matrix representation in the Krylov basis or the CMV basis (see Appendix \ref{App: D} for the latter). For unitary evolution represented in the Krylov operator basis \eqref{Eq: Krylov basis K-matrix}, one obtains the upper Hessenberg form \eqref{Eq: upper-Hessenberg} using $\langle P_j(z), P_{k+1}(z) \rangle = \delta_{j,k+1}$ and \eqref{Eq: P P^* inner product}.

\section{CMV basis}
\label{App: D}
Here, we will show that the CMV basis \eqref{Eq: CMV basis} leads to the same matrix elements as the ITFIM discussed in Section \ref{Sec: ITFIM and CMV basis}. First, we check whether this basis is orthonormal. Note that the CMV basis is written as a Laurent series, i.e., it contains negative powers of $z$ in contrast to the OPUC. In order to apply the orthogonal property of the OPUC, one has to apply $z^{-1} = \overline{z}$ in the inner product so that the expression can be rearranged as inner product between polynomials. For $\langle \chi_{2k}, \chi_{2l} \rangle$ with $k \geq l$,
\begin{align}
    \langle \chi_{2k}, \chi_{2l} \rangle &= \langle z^{-k}P_{2k}, z^{-l}P_{2l} \rangle \nonumber\\
    &= \langle P_{2k}, z^{k-l}P_{2l} \rangle =\delta_{kl},
\end{align}
where in the last line we use the fact that $P_{2k}$ is orthogonal to $\{1, z, \ldots, z^{2k-1} \}$ and $z^{k-l}P_{2l}$ consists of terms like $\{ z^{k-l},z^{k-l+1},\ldots, z^{k+l} \}$. Therefore, the inner product is not zero only when $k = l$. By a similar argument, for $k \leq l$, we obtain
\begin{align}
    \langle \chi_{2k}, \chi_{2l} \rangle = \langle z^{l-k}P_{2k}, P_{2l} \rangle = \delta_{kl}.
\end{align}
Next, we check $\langle \chi_{2k+1}, \chi_{2l+1} \rangle$ with $k \geq l$,
\begin{align}
    \langle \chi_{2k+1}, \chi_{2l+1} \rangle &= \langle z^{-k-1}P_{2k+1}^*, z^{-l-1}P_{2l+1}^* \rangle \nonumber\\
    &= \langle P_{2k+1}^*, z^{k-l}P_{2l+1}^* \rangle = \delta_{kl},
\end{align}
where in the last line we apply the fact that $P_{2k+1}^*$ is orthogonal to $\{z,z^2,\ldots, z^{2k+1} \}$ and $z^{k-l}P_{2l+1}^*$ consists of terms like $\{ z^{k-l},z^{k-l+1},\ldots, z^{k+l+1} \}$. Therefore, the inner product is not zero only when $k = l$. By a similar argument, for $k \leq l$,
\begin{align}
    \langle \chi_{2k+1}, \chi_{2l+1} \rangle = \langle z^{l-k}P_{2k+1}^*, P_{2l+1}^* \rangle = \delta_{kl}.
\end{align}
Finally, we check $\langle \chi_{2k}, \chi_{2l+1}  \rangle$ with $k > l$, 
\begin{align}
    \langle \chi_{2k}, \chi_{2l+1}  \rangle &= \langle z^{-k} P_{2k}, z^{-l-1}P_{2l+1}^* \rangle \nonumber\\
    &= \langle  P_{2k}, z^{k-l-1}P_{2l+1}^* \rangle = 0,
\end{align}
where in the last line we apply the fact that $P_{2k}$ is orthogonal to $\{1,z,\ldots, z^{2k-1} \}$ and $z^{k-l-1}P_{2l+1}^*$ consists of terms like $\{ z^{k-l-1},z^{k-l},\ldots, z^{k+l} \}$. The inner product is zero since $ 2k -1 \geq k+l$ for $k > l$. By a similar argument, for $k \leq l$,
\begin{align}
    \langle \chi_{2k}, \chi_{2l+1}  \rangle = \langle  z^{l-k+1}P_{2k}, P_{2l+1}^* \rangle = 0,
\end{align}
where in the last line we apply the fact that $P_{2l+1}^*$ is orthogonal to $\{z,z^2,\ldots, z^{2l+1} \}$ and $z^{l-k+1}P_{2k}$ consists of terms like $\{ z^{l-k+1},z^{l-k+2},\ldots, z^{l+k+1} \}$. The inner product is zero since $ 2l +1 \geq l+k+1$ for $k \leq l$. Therefore, we have shown that the CMV basis is orthonormal.

Now, we are ready to compute the matrix element of $z$ in the CMV basis. Let us compute \eqref{Eq: CMV basis matrix element} one by one. For $\langle \chi_{2k}, z \chi_{2k-2} \rangle$, we first rearrange the expression as
\begin{align}
    \langle \chi_{2k}, z \chi_{2k-2} \rangle &= \langle z^{-k}P_{2k}, zz^{-k+1}P_{2k-2} \rangle \nonumber\\
    &=\langle P_{2k}, z^2P_{2k-2} \rangle.
\end{align}
We first apply \eqref{Eq: Szego P-1} to $z^2P_{2k-2}$,
\begin{align}
    \langle \chi_{2k}, z \chi_{2k-2} \rangle &= \langle P_{2k}, z\left(\rho_{2k-2}P_{2k-1} + \overline{\alpha_{2k-2}}P_{2k-2}^* \right) \rangle\nonumber \\
    &= \langle P_{2k},  \rho_{2k-2}zP_{2k-1} \rangle,
\end{align}
where $\langle P_{2k}, z P_{2k-2}^* \rangle = 0$ is used in the second line. We apply \eqref{Eq: Szego P-1} again,
\begin{align}
    \langle \chi_{2k}, z \chi_{2k-2} \rangle &= \langle P_{2k},  \rho_{2k-2}\left( \rho_{2k-1}P_{2k} + \overline{\alpha_{2k-1}}P_{2k-1}^* \right) \rangle \nonumber\\
    &= \langle P_{2k},  \rho_{2k-2}\rho_{2k-1}P_{2k} \rangle
    \nonumber\\ &= \rho_{2k-2}\rho_{2k-1},
\end{align}
where $\langle P_{2k}, z P_{2k-2}^* \rangle = 0$ is used in the second line.

For $\langle \chi_{2k}, z \chi_{2k-1} \rangle$,  
\begin{align}
    \langle \chi_{2k}, z \chi_{2k-1} \rangle &= \langle z^{-k}P_{2k}, zz^{-k}P_{2k-1}^* \rangle \nonumber\\
    &= \langle P_{2k}, zP_{2k-1}^* \rangle.
\end{align}
Next, we apply the inverse Szeg\H{o} recurrence relation \eqref{Eq: inverse Szego P-2},
\begin{align}
    \langle \chi_{2k}, z \chi_{2k-1} \rangle &= \langle P_{2k}, z\left(-\alpha_{2k-2}P_{2k-1} + \rho_{2k-2} P_{2k-2}^*\right) \rangle \nonumber\\
    &=\langle P_{2k}, -z\alpha_{2k-2}P_{2k-1} \rangle\nonumber\\
    &= -\alpha_{2k-2}\rho_{2k-1},
\end{align}
where $\langle P_{2k}, zP^*_{2k-2} \rangle = 0$ is used in the second line. In addition, we have used  the identity \eqref{Eq: K-matrix in P} and the lower off-diagonal result of \eqref{Eq: upper-Hessenberg} in the third line.

For $\langle \chi_{2k}, z \chi_{2k} \rangle$,
\begin{align}
    \langle \chi_{2k}, z \chi_{2k} \rangle = \langle z^{-k}P_{2k}, zz^{-k}P_{2k} \rangle = \langle P_{2k}, zP_{2k} \rangle.
\end{align}
We apply \eqref{Eq: Szego P-1} and obtain
\begin{align}
    \langle \chi_{2k}, z \chi_{2k} \rangle &= \langle P_{2k}, \rho_{2k}P_{2k+1} + \overline{\alpha_{2k}}P_{2k}^* \rangle \nonumber\\
    &= \langle P_{2k},\overline{\alpha_{2k}}P_{2k}^* \rangle\nonumber\\
    &= -\alpha_{2k-1}\overline{\alpha_{2k}},
\end{align}
where $\langle P_{2k}, P_{2k+1} \rangle = 0$ and \eqref{Eq: P P^* inner product} are used in second and third line respectively.

For $\langle \chi_{2k}, z \chi_{2k+1} \rangle$, it is straightforward to show that
\begin{align}
    \langle \chi_{2k}, z \chi_{2k+1} \rangle &= \langle z^{-k}P_{2k}, zz^{-k-1}P_{2k+1}^* \rangle \nonumber\\
    &= \langle P_{2k}, P_{2k+1}^* \rangle \nonumber\\
    &= -\alpha_{2k-1}\rho_{2k},
\end{align}
where we have applied \eqref{Eq: P P^* inner product} in the third line.

For $\langle \chi_{2k+1}, z \chi_{2k} \rangle$,
\begin{align}
    \langle \chi_{2k+1}, z \chi_{2k} \rangle &= \langle z^{-k-1}P_{2k+1}^*, zz^{-k}P_{2k} \rangle\nonumber\\
    &= \langle P_{2k+1}^*, z^2 P_{2k} \rangle.
\end{align}
From \eqref{Eq: Szego P-1}, we have
\begin{align}
    \langle \chi_{2k+1}, z \chi_{2k} \rangle & = \langle P_{2k+1}^*, z\left(\rho_{2k}P_{2k+1} + \overline{\alpha_{2k}}P_{2k}^* \right)\rangle\nonumber\\
    &= \langle P_{2k+1}^*, \rho_{2k}zP_{2k+1} \rangle,
\end{align}
where $\langle P^*_{2k+1}, zP^*_{2k} \rangle = 0$ is applied. Using \eqref{Eq: Szego P-1} again, we obtain
\begin{align}
    \langle \chi_{2k+1}, z \chi_{2k} \rangle &= \langle P_{2k+1}^*, \rho_{2k}\left( \rho_{2k+1}P_{2k+2} + \overline{\alpha_{2k+1}}P_{2k+1}^* \right) \rangle \nonumber\\
    &=  \langle P_{2k+1}^*, \rho_{2k}\overline{\alpha_{2k+1}}P_{2k+1}^* \rangle\nonumber\\
    &= \rho_{2k}\overline{\alpha_{2k+1}},
\end{align}
where $\langle P_{2k+1}^*, P_{2k+2}  \rangle = 0$ is used in the second line.

For $\langle \chi_{2k+1}, z \chi_{2k+1} \rangle$,
\begin{align}
    \langle \chi_{2k+1}, z \chi_{2k+1} \rangle &= \langle z^{-k-1}P_{2k+1}^*, zz^{-k-1}P_{2k+1}^* \rangle \nonumber\\
    &= \langle P_{2k+1}^*, zP_{2k+1}^* \rangle.
\end{align}
Using \eqref{Eq: inverse Szego P-2}, 
\begin{align}
    \langle \chi_{2k+1}, z \chi_{2k+1} \rangle &= \langle P_{2k+1}^*, z\left(\rho_{2k}P_{2k}^* - \alpha_{2k}P_{2k+1}\right) \rangle\nonumber\\
    &= \langle P_{2k+1}^*, - z\alpha_{2k}P_{2k+1} \rangle,
\end{align}
where $\langle P^*_{2k+1}, zP^*_{2k} \rangle = 0$ is applied. By \eqref{Eq: Szego P-1},
\begin{align}
    &\langle \chi_{2k+1}, z \chi_{2k+1} \rangle \nonumber\\
    &= \langle P_{2k+1}^*, - \alpha_{2k}\left( \rho_{2k+1}P_{2k+2} + \overline{\alpha_{2k+1}}P_{2k+1}^* \right) \rangle \nonumber\\
    &= -\alpha_{2k}\overline{\alpha_{2k+1}},
\end{align}
where $\langle P_{2k+1}^*, P_{2k+2} \rangle = 0$ is used.

For $\langle \chi_{2k+1}, z \chi_{2k+2} \rangle$, we obtain
\begin{align}
    \langle \chi_{2k+1}, z \chi_{2k+2} \rangle &= \langle z^{-k-1}P_{2k+1}^*, zz^{-k-1}P_{2k+2} \rangle\nonumber\\
    &= \langle P_{2k+1}^*, zP_{2k+2} \rangle.
\end{align}
From \eqref{Eq: Szego P-1}, we obtain
\begin{align}
    \langle \chi_{2k+1}, z \chi_{2k+2} \rangle &= \langle P_{2k+1}^*, \left( \rho_{2k+2}P_{2k+3} + \overline{\alpha_{2k+2}}P_{2k+2}^* \right) \rangle \nonumber\\
    &= \langle P_{2k+1}^*,  \overline{\alpha_{2k+2}}P_{2k+2}^*  \rangle\nonumber\\
    &= \rho_{2k+1}\overline{\alpha_{2k+2}},
\end{align}
where $\langle P_{2k+1}^*, P_{2k+3} \rangle = 0$ and \eqref{Eq: P* P* inner product} are used.

For the last term $\langle \chi_{2k+1}, z \chi_{2k+3} \rangle$, we obtain
\begin{align}
    \langle \chi_{2k+1}, z \chi_{2k+3} \rangle &= \langle z^{-k-1}P_{2k+1}^*, zz^{-k-2}P_{2k+3}^* \rangle \nonumber\\
    &= \langle P_{2k+1}^*, P_{2k+3}^* \rangle \nonumber\\
    &= \rho_{2k+1}\rho_{2k+2},
\end{align}
where \eqref{Eq: P* P* inner product} is applied.

In summary, we have derived the results in \eqref{Eq: CMV basis matrix element}. All other matrix elements are zero as can be checked by computations similar to those performed above.   

\section{Construction of a lower degree OPUC from a given $P_k(z)$}
\label{App: E}

If $P_k(z)$ is given, one can construct the lower degree OPUC from the monic Szeg\H{o} recurrence relations \eqref{Eq: monic Szego-1} as follows. Since $P_k(z)$ is known, one can first construct the monic polynomial $p_k(z)$ by normalizing the prefactor of the $z^n$ term in $P_k(z)$. Evaluating \eqref{Eq: monic Szego-1} at $z=0$, one obtains
\begin{align}
    p_{k}(0) = 0\times p_{k-1}(0) - \overline{\alpha_{k-1}} p_{k-1}^*(0) = -\overline{\alpha_{k-1}}.
\end{align}
The above shows that $P_k(z)$ uniquely determines the Verblunsky coefficient $\alpha_{k-1}$ (or equivalently, the Krylov angle $\theta_{k-1}$ for real $\alpha_{k-1}$). Next, from the inverse relation \eqref{Eq: monic inverse p-1} one obtains
\begin{align}
    p_{k-1}(z) = \frac{1}{\rho_{k-1}^2 z} \left[ p_{k}(z) + \overline{\alpha_{k-1}}p_{k}^*(z) \right].
\end{align}
Note that everything on the right hand side is uniquely determined from a given  $P_k(z)$ as discussed above. Therefore, $p_{k-1}(z)$ is uniquely determined. The other monic polynomials $\{ p_0, p_1, \ldots, p_k \}$ and Verblunsky coefficients $\{ \alpha_0, \alpha_1, \ldots, \alpha_{k-1} \}$ are constructed by repeating this procedure. Finally, we can compute $\{\kappa_{11}, \kappa_{22}, \ldots, \kappa_{kk} \}$ from \eqref{Eq: kappa rho}. Combining the known monic polynomials $\{ p_0, p_1, \ldots, p_k \}$ and $\{\kappa_{11}, \kappa_{22}, \ldots, \kappa_{kk} \}$, we obtain the OPUC $\{ P_0, P_1, \ldots, P_k \}$ according to \eqref{Eq: monic p}.
\section{Bernstein–Szeg\H{o} approximation for finite Krylov space dimension}
\label{App: F}
We demonstrate that the Bernstein–Szeg\H{o} approximation reproduces the exact spectral function when the Krylov space has a finite dimension. We consider the simplest example, $\alpha_0 = 1$, where the autocorrelation function is constant, $A(n) = 1$ and the Krylov space has dimension one. Therefore, one expects to see a single delta function at $\omega = 0$.
From the Szeg\H{o} recurrence relation \eqref{Eq: Szego P-1}, we obtain the analytic expression for $P_1(z)$, 
\begin{align}
    P_1(z) = \frac{z-\overline{\alpha_0}}{\rho_0},
\end{align}
where the initial conditions $P_0 = P_0^* = 1$ are applied. Note that this expression diverges when $\rho_0 = 0$, which occurs for $\alpha_0 = 1$. To regularize the expression, we set $\alpha_0 = e^{-\epsilon}$ with $\epsilon > 0$. The corresponding Bernstein–Szeg\H{o} approximation becomes
\begin{align}
    \Phi_1(e^{i\omega}) &= \frac{1}{2\pi}\frac{1}{|P_1(e^{i\omega})|^2}= \frac{1}{2\pi}\frac{1-e^{-2\epsilon}}{(e^{i\omega} - e^{-\epsilon})(e^{-i\omega} - e^{-\epsilon})}.
\end{align}
To connect this result to the delta function, we first rearrange the denominator as,
\begin{align}
    &\frac{1}{2\pi}\frac{1-e^{-2\epsilon}}{(e^{i\omega} - e^{-\epsilon})(e^{-i\omega} - e^{-\epsilon})} \nonumber\\
    &= \frac{1}{2\pi} \frac{1-e^{-2\epsilon}}{(1 - e^{-\epsilon-i\omega})(1 - e^{-\epsilon+i\omega})}.
\end{align}
The fraction can then be decomposed into a sum of three terms,
\begin{align}
    &\frac{1}{2\pi} \frac{1-e^{-2\epsilon}}{(1 - e^{-\epsilon-i\omega})(1 - e^{-\epsilon+i\omega})}\nonumber\\
    &=\frac{1}{2\pi}\left( \frac{1}{1-e^{-\epsilon+i\omega}} + \frac{1}{1-e^{-\epsilon-i\omega}}-1\right).
\end{align}
According to the Floquet  Sokhotski-Plemelj theorem, see Appendix \ref{App: I} for the derivation,
\begin{align}
    \lim_{\epsilon\rightarrow 0^+}\frac{1}{1-e^{-\epsilon -i\omega}} = \frac{1}{2} - \frac{i}{2}\cot\frac{\omega}{2} + \pi \delta_F(\omega).
\end{align}
Taking the limit $\epsilon \rightarrow 0^+$, the exact spectral function is recovered as
\begin{align}
    \Phi(e^{i\omega}) = \lim_{\epsilon\rightarrow 0^+} \frac{1}{2\pi}\frac{1}{|P_1(e^{i\omega})|^2}\Big|_{\alpha_0 = e^{-\epsilon}} = \delta_{F}(\omega).
\end{align}
In general, when $|\alpha_k^*| = 1$, the Krylov space has a finite dimension  of $k^*+1$. Thus there are $k^*+1$ eigenvalues, and therefore the spectral function consists of $k^*+1$ delta-function peaks.

\section{Relation between the continued fraction expansion of the Laplace transformation, the OPUC, and the Szeg\H{o} recurrence formula}
\label{App: G}

We first prove the relation \eqref{Eq: Bernstein approx truncated continued fraction}. According to \eqref{Eq: Bernstein large k}, \eqref{Eq: spectral Laplace} and \eqref{Eq: Laplace Continued fraction}, we conclude the following relation
\begin{align}
    &\lim_{k\rightarrow \infty}  \frac{1}{|P_k(e^{i\omega})|^2}\nonumber\\
    &= -1 + 2\lim_{\epsilon\rightarrow 0^+} \text{Re}\left[\underset{M\rightarrow \infty}{\lim}G_C(e^{\epsilon+i\omega},\theta; M)\right],\ \forall\ \epsilon>0
\end{align}
If $\theta_{j \geq \Tilde{k}} = \pi/2$ for some integer $\Tilde{k}$, the left hand side becomes 
\begin{align}
    \lim_{k\rightarrow \infty}  \frac{1}{|P_k(e^{i\omega})|^2} = \frac{1}{|P_{\Tilde{k}-1}(e^{i\omega})|^2},\ \forall\  \theta_{k\geq \Tilde{k}} = \pi/2.
    \label{Eq: k tilde-1}
\end{align}
Above we have used that $\theta_{k\geq \Tilde{k}} = \pi/2$ which is equivalent to
$\alpha_{k\geq \Tilde{k}-1} =0$. Thus from
\eqref{Eq: Szego P-1} it follows that
$P_{k\geq  \Tilde{k}-1}(z) = z^{k-\Tilde{k}+1}P_{\Tilde{k}-1}(z)$. 
By setting $z=e^{i\omega}$, one obtains $|P_k(e^{i\omega})|^2 = |P_{\Tilde{k}-1}(e^{i\omega})|^2$ for $k \geq \Tilde{k}-1$.

The continued fraction $G_C$ can be explicitly expressed as 
\begin{align}
    \underset{M\rightarrow \infty}{\lim} G_C(z,\theta;M)
    &=  \frac{a_0}{b_0 + \frac{a_1}{\frac{\ddots}{b_{\Tilde{k}-2} + \frac{a_{\Tilde{k}-1}}{b_{\Tilde{k}-1}+X}}}},\ \forall\ |z| >1, 
\end{align}
where
\begin{align}
    X &= a_{\Tilde{k}}/a_{\Tilde{k}+1}/a_{\Tilde{k}+2}/\ldots,\ \forall\ \theta_{j\geq \Tilde{k}} = \pi/2,\ |z| >1\nonumber\\
    &=  \Big\{\begin{array}{cc} 1/z^2/1/z^2/\ldots =
    0 & \text{if}\ \Tilde{k}\ \text{is odd};\\
    z^2/1/z^2/1/\ldots =\text{divergent} & \text{else}.
    \end{array}
\end{align}
The above limit of $X$ is derived from the maximally ergodic case where all Krylov angles are set to $\pi/2$ \cite{yeh2025moment}. For this case, the autocorrelation is given by $A(n) = \delta_{0n}$ and its Laplace transformation is simply $G_L(z) = 1$. The corresponding continued fraction expansion is $G_C(z) = z/(z+X)$ with $X = 1/z^2/1/z^2/\ldots$. By matching $G_L(z) = G_C(z)$ for this maximally ergodic example, one concludes $1/z^2/1/z^2/\ldots = 0$, arriving at the result for odd $\Tilde{k}$ above. The even $\Tilde{k}$ case can be thought as the inverse of the odd $\Tilde{k}$ case and multiplied by $z^2$, yielding a divergent result.

For $\Tilde{k}$ being an odd number with $\theta_{j\geq \Tilde{k}} = \pi/2$ and $X=0$, the continued fraction expansion is simplified to
\begin{align}
    \underset{M\rightarrow \infty}{\lim} G_C(z,\theta;M)
    &=  \frac{a_0}{b_0 + \frac{a_1}{\frac{\ddots}{b_{\Tilde{k}-2} + \frac{a_{\Tilde{k}-1}}{b_{\Tilde{k}-1}}}}},\ \forall\ |z| >1\nonumber\\
    &=G_C(z,\theta;\Tilde{k}-1)\Big|_{\theta_{\Tilde{k}}=\pi/2},\ \forall\ |z| >1,
    \label{Eq: k tilde-2}
\end{align}
where $b_{\Tilde{k}-1}$ contains $\theta_{\Tilde{k}-1}$ and $\theta_{\Tilde{k}}$ (see \eqref{Eq: a b coefficient}) so that we have to set $\theta_{\Tilde{k}} = \pi/2$ in the last line. For even $\Tilde{k}$ with $\theta_{j\geq \Tilde{k}} = \pi/2$, since $X$  is divergent, we have
\begin{align}
    \underset{M\rightarrow \infty}{\lim} G_C(z,\theta;M)
    &=  \frac{a_0}{b_0 + \frac{a_1}{\frac{\ddots}{b_{\Tilde{k}-3} + \frac{a_{\Tilde{k}-2}}{b_{\Tilde{k}-2}}}}},\ \forall\ |z| >1\nonumber\\
    &=G_C(z,\theta;\Tilde{k}-2),\ \forall\ |z| >1,
    \label{Eq: k tilde-3}
\end{align}
where $b_{\Tilde{k}-2}$ does not contain $\theta_{\Tilde{k}}$ and therefore there are no constraints in the last line. By setting $\tilde{k} = 2k+1$, we can derive \eqref{Eq: P GC-1} from \eqref{Eq: k tilde-1} and \eqref{Eq: k tilde-2}. \eqref{Eq: P GC-2} is obtained from \eqref{Eq: k tilde-1} and \eqref{Eq: k tilde-3} by setting $\tilde{k} = 2k+2$.

Next, we show how the monic polynomials in the Szeg\H{o} recurrence relations \eqref{Eq: monic Szego} appear naturally in the recurrence relations for the continued fraction expansion \eqref{Eq: f h recurrence}. We start with \eqref{Eq: monic Szego-1} and consider $p_{2k+1}$,
\begin{align}
    p_{2k+1} = zp_{2k} - \overline{\alpha_{2k}}p_{2k}^*,
\end{align}
where we omit the $z$-dependence of monic polynomials for conciseness. Then we apply \eqref{Eq: monic Szego-1} again on $p_{2k}$,
\begin{align}
    p_{2k+1} 
    =z (zp_{2k-1} - \overline{\alpha_{2k-1}}p_{2k-1}^*) - \overline{\alpha_{2k}}p_{2k}^*.
\end{align}
Finally, we apply \eqref{Eq: monic Szego-2} on $p^*_{2k-1}$,
\begin{align}
    &p_{2k+1}\nonumber\\
    &=z \left[zp_{2k-1} - \overline{\alpha_{2k-1}}(p_{2k}^* + \alpha_{2k-1}zp_{2k-1}) \right] - \overline{\alpha_{2k}}p_{2k}^*\nonumber\\
    &= (-z\overline{\alpha_{2k-1}} - \overline{\alpha_{2k}})p_{2k}^* + z^2\rho_{2k-1}^2 p_{2k-1}.
\end{align}
By using $\alpha_{k} = (-1)^k \cos\theta_{k+1}$ and $\rho_{k} = \sin\theta_{k+1}$ for real Verblunsky coefficients, we arrive at
\begin{align}
    &p_{2k+1} \nonumber\\
    &= (z\cos\theta_{2k} - \cos\theta_{2k+1})p_{2k}^* +z^2\sin^2\theta_{2k} p_{2k-1}\nonumber\\
    &= b_{2k}p_{2k}^* + a_{2k}p_{2k-1},
\end{align}
where \eqref{Eq: a b coefficient} is applied in the last line. For the recurrence relation with $a_{2k-1}$ and $b_{2k-1}$, we start with $p_{2k}^*$ and follow similar steps as above
\begin{align}
    &p_{2k}^*\nonumber\\
    &= p_{2k-1}^* - \alpha_{2k-1}zp_{2k-1}\nonumber\\
    &= (p_{2k-2}^* - \alpha_{2k-2}zp_{2k-2}) - \alpha_{2k-1}zp_{2k-1}\nonumber\\
    &=\left[ p_{2k-2}^* - \alpha_{2k-2}(p_{2k-1} + \overline{\alpha_{2k-2}}p_{2k-2}^*) \right] - \alpha_{2k-1}zp_{2k-1}\nonumber\\
    &= (-z\alpha_{2k-1} - \alpha_{2k-2})p_{2k-1} +\rho_{2k-2}^2p_{2k-2}^*\nonumber\\
    &=b_{2k-1}p_{2k-1} + a_{2k-1}p_{2k-2}^*,
\end{align}
where \eqref{Eq: monic Szego-2} is applied in the first and second lines. \eqref{Eq: monic Szego-1} and \eqref{Eq: a b coefficient} are applied in the third and last lines.

In summary, for real Verblunsky coefficients, the monic polynomials satisfy
\begin{align}
   &p_{2k+1} = b_{2k}p_{2k}^* + a_{2k}p_{2k-1};\\
   &p_{2k}^* = b_{2k-1}p_{2k-1} + a_{2k-1}p_{2k-2}^*.
\end{align}
These are the same recurrence relations as \eqref{Eq: f h recurrence}. Now we have to also match initial conditions. One can explicitly check that  $h_{-1} = p_0^*=1$, $h_{0} = z-\cos\theta_1=p_1$ and therefore
\begin{align}
    h_{2k-1} = p_{2k}^*,\ h_{2k} = p_{2k+1}.
\end{align}
However, $f_k$ is constructed from a different set of initial conditions: $f_{-2} = 1, f_{-1}=0$ than $p_{k},p^*_{k}$, so that they are not as simply related as the $h_k$.

\section{Analytic expressions for the OPUC for $m$-persistent autocorrelation functions with $m=1,2,3,4,6$}
\label{App: H}
We start with the $m=1$ case, where the Krylov angles are given by \eqref{Eq: m=1 Krylov angle}. By explicitly writing down the first few monic polynomials via \eqref{Eq: monic Szego} with \eqref{Eq: Verblunsky Krylov angle}, one gets
\begin{align}
    &p_1(z) = z-A,\\
    &p_2(z) = z^2 - \frac{A}{1+A}(z+1),\\
    &p_3(z) = z^3 - \frac{A}{1+2A}(z^2+z+1).
\end{align}
This pattern persists and the general form is
\begin{align}
    p_k(z) &= z^{k} - \frac{A}{1+(1-k)A}\sum_{j=0}^{k-1}z^j\nonumber\\
    &= z^{k} - (-1)^{k-1}\cos\theta^{(1)}_k\sum_{j=0}^{k-1}z^j
\end{align}
Using \eqref{Eq: monic p} and \eqref{Eq: kappa rho}, one recovers the result in \eqref{Eq: P m=1}. This result is called the single inserted mass point in Ref.~\cite{simon2005orthogonalpart1} since its spectral function \eqref{Eq: spectral function m=1} has one single peak.

The OPUC for $m=2$ can be derived by manipulating the Szeg\H{o} recurrence relations for $m=1$,
\begin{subequations}
\label{Eq: Szego relation m=1}
\begin{align}
    &\sin\theta^{(1)}_{k+1} P^{(1)}_{k+1}(z) = zP^{(1)}_k(z) - (-1)^k \cos\theta^{(1)}_{k+1}P^{(1)*}_k(z);\\
    &\sin\theta^{(1)}_{k+1} P^{(1)*}_{k+1}(z) = P^{(1)*}_k(z) - (-1)^k \cos\theta^{(1)}_{k+1} z P^{(1)}_k(z).
\end{align}
\end{subequations}
Above, the relations between the Verblunsky coefficients and the Krylov angles \eqref{Eq: Verblunsky Krylov angle} were used. The Krylov angles for $m = 2$ are related to those for $m=1$ according to \eqref{Eq: m=2 Krylov angle}:  $\sin\theta^{(1)}_{k} = \sin\theta^{(2)}_{k}$ and $\cos\theta^{(1)}_{k}=(-1)^{k}\cos\theta^{(2)}_{k}, k\geq 1$. \eqref{Eq: Szego relation m=1} becomes
\begin{subequations}
\begin{align}
    &\sin\theta^{(2)}_{k+1} P^{(1)}_{k+1}(z) = zP^{(1)}_k(z) + \cos\theta^{(2)}_{k+1}P^{(1)*}_k(z);\\
    &\sin\theta^{(2)}_{k+1} P^{(1)*}_{k+1}(z) = P^{(1)*}_k(z) + \cos\theta^{(2)}_{k+1} z P^{(1)}_k(z).
\end{align}
\end{subequations}
Then we flip the sign of $z$ in both equations $z \rightarrow -z$ and multiply the first one by $(-1)^{k+1}$, giving
\begin{subequations}
\begin{align}
    &\sin\theta^{(2)}_{k+1}(-1)^{k+1}P^{(1)}_{k+1}(-z)\nonumber\\
    &= z(-1)^{k}P^{(1)}_k(-z) - (-1)^{k} \cos\theta^{(2)}_{k+1}P^{(1)*}_k(-z);\\
    &\sin\theta^{(2)}_{k+1} P^{(1)*}_{k+1}(-z)\nonumber\\
    &= P^{(1)*}_k(-z) -(-1)^{k} \cos\theta^{(2)}_{k+1} z (-1)^k P^{(1)}_k(-z),
\end{align}
\end{subequations}
where in the last line we apply $1 = (-1)^{2k} = (-1)^k (-1)^k$. By identifying $P^{(2)}_k(z) = (-1)^kP^{(1)}_k(-z)$ and $P^{(2)*}_k(z) = P^{(1)*}_k(-z)$, the above equations become the Szeg\H{o} recurrence relations for $m=2$,
\begin{subequations}
\begin{align}
    &\sin\theta^{(2)}_{k+1} P^{(2)}_{k+1}(z) = zP^{(2)}_k(z) - (-1)^k \cos\theta^{(2)}_{k+1}P^{(2)*}_k(z);\\
    &\sin\theta^{(2)}_{k+1} P^{(2)*}_{k+1}(z) = P^{(2)*}_k(z) - (-1)^k \cos\theta^{(2)}_{k+1} z P^{(2)}_k(z).
\end{align}
\end{subequations}
Therefore, OPUC relation \eqref{Eq: P m=2} is consistent with the Krylov angles for $m=2$ \eqref{Eq: m=2 Krylov angle}.

For the $m=4$ case, we start from the Szeg\H{o} recurrence relations,
\begin{align}
    &P^{(4)}_{2k+1}(z) = z P^{(4)}_{2k}(z),\ P^{(4)*}_{2k+1}(z) = P^{(4){*}}_{2k}(z),\\
    &\sin\theta^{(1)}_{k+1} P^{(4)}_{2k+2}(z) = zP^{(4)}_{2k+1}(z) + \cos\theta^{(1)}_{k+1}P^{(4)*}_{2k+1}(z),\\
    &\sin\theta^{(1)}_{k+1} P^{(4)*}_{2k+2}(z) = P^{(4)*}_{2k+1}(z) + \cos\theta^{(1)}_{k+1} z P^{(4)}_{2k+1}(z),
\end{align}
where the Krylov angle relation \eqref{Eq: m=4 Krylov angle} has been used. By combining the first line with the second and the third one, we obtain
\begin{align}
    &\sin\theta^{(1)}_{k+1} P^{(4)}_{2k+2}(z) = z^2P^{(4)}_{2k}(z) + \cos\theta^{(1)}_{k+1}P^{(4)*}_{2k}(z);\\
    &\sin\theta^{(1)}_{k+1} P^{(4)*}_{2k+2}(z) = P^{(4)*}_{2k}(z) + \cos\theta^{(1)}_{k} z^2 P^{(4)}_{2k}(z).
\end{align}
Then, multiplying the first line by $(-1)^{k+1}$ and applying the identity, $z^2 = -(-z^2)(-1)^{k}(-1)^{k}$ in the last term of the second line, we obtain 
\begin{align}
    &\sin\theta^{(1)}_{k+1} (-1)^{k+1} P^{(4)}_{2k+2}(z)\nonumber\\
    &= -z^2 (-1)^k P^{(4)}_{2k}(z) - (-1)^k \cos\theta^{(1)}_{k+1}P^{(4)*}_{2k}(z);\\
    &\sin\theta^{(1)}_{k+1} P^{(4)*}_{2k+2}(z)\nonumber\\
    &= P^{(4)*}_{2k}(z) - (-1)^{k}\cos\theta^{(1)}_{k+1} (-z^2) (-1)^{k}P^{(4)}_{2k}(z).
\end{align}
Finally, we identify $(-1)^k P^{(4)}_{2k}(z) = P^{(1)}_k(-z^2)$ and $P^{(4)^*}_{2k}(z)=P^{(1)*}_k(-z^2)$ and the equations become
\begin{align}
    &\sin\theta^{(1)}_{k+1} P^{(1)}_{k+1}(-z^2)\nonumber\\
    &= -z^2 P^{(1)}_{k}(-z^2) - (-1)^k \cos\theta^{(1)}_{k+1}P^{(1)*}_k(-z^2);\\
    &\sin\theta^{(1)}_{k+1} P^{(1)*}_{k+1}(-z^2)\nonumber\\
    &= P^{(1)*}_{k}(-z^2) - (-1)^{k}\cos\theta^{(1)}_{k+1} (-z^2) P^{(1)}_{k}(-z^2).
\end{align}
The above results can be obtained from \eqref{Eq: Szego relation m=1} by taking $z \rightarrow -z^2$. Therefore, OPUC relation \eqref{Eq: P m=4} is consistent with Krylov angles for $m=4$ \eqref{Eq: m=4 Krylov angle}.  

There is no direct relation between $m=3$ and $m=1$, as far as we know. We explicitly write down its first few monic polynomials using \eqref{Eq: monic Szego} and \eqref{Eq: m=3 Krylov angle}. We first consider $p_{3k-2}(z)$,
\begin{widetext}
\begin{align}
    &p_1^{(3)}(z) = z + \frac{A}{2};\\
    &p_4^{(3)}(z) = z^4 + \frac{A}{2+3A}(z^3 + 1) + \frac{A(2+4A)}{4+8A+3A^2}z^2 - \frac{A(4+5A)}{4+8A+3A^2}z; \\
    &p_7^{(3)}(z) = z^7 + \frac{A}{2+6A}(z^6 + z^3 +1) + \frac{A(2+7A)}{4+20A+24A^2}(z^5 + z^2) - \frac{A(4+11A)}{4+20A+24A^2}(z^4 + z).
\end{align}
In general, it has the following form
\begin{align}
    p^{(3)}_{3k-2}(z) &= z^{3k-2} + \frac{A}{2+(3k-3)A}\left( \sum_{j=0}^{k-1} z^{3j} \right)  + \frac{A[2+(3k-2)A]}{[2+(3k-5)A][2+(3k-3)A]}\left( \sum_{j=1}^{k-1}z^{3j-1} \right)\nonumber\\
   &- \frac{A[4+(6k-7)A]}{[2+(3k-5)A][2+(3k-3)A]}\left( \sum_{j=1}^{k-1}z^{3j-2}  \right)\nonumber\\
    &= z^{3k-2} + (-1)^{k}\cos\theta^{(3)}_{3k-2}\left( \sum_{j=0}^{k-1} z^{3j} \right) + (-1)^{k-1}\frac{\cos\theta^{(3)}_{3k-3}\cos\theta^{(3)}_{3k-2}}{\cos\theta^{(3)}_{3k}}\left( \sum_{j=1}^{k-1}z^{3j-1}\right)\nonumber\\
    & +(-1)^k \frac{\cos\theta^{(3)}_{3k-3}}{2\cos\theta^{(3)}_{3k-1}}\left( \cos\theta^{(3)}_{3k-2} - \cos\theta^{(3)}_{3k-1} \right)\left( \sum_{j=1}^{k-1}z^{3j-2} \right).
\end{align}
In a similar manner,
$p_{3k-1}(z)$ and $p_{3k}(z)$,
\begin{align}
    &p^{(3)}_2(z) = z^2 + \frac{A}{2-A}(z+1),\\
    &p^{(3)}_3(z) = z^3 + \frac{A}{2+A}(z^2+z) -\frac{2A}{2+A},\\
    &p^{(3)}_5(z) = z^5 + \frac{A}{2+2A}(z^4 + z^3 + z + 1) - \frac{2A}{2+2A}z^2,\\
    &p^{(3)}_6(z) = z^6 + \frac{A}{2+4A}(z^5+z^4+z^2+z) - \frac{2A}{2+4A}(z^3+1),\\
    &p^{(3)}_8(z) = z^8 + \frac{A}{2+5A}(z^7+z^6+z^4+z^3+z+1) - \frac{2A}{2+5A}(z^5+z^2),\\
    &p^{(3)}_9(z) = z^9 + \frac{A}{5+7A}(z^8+z^7+z^5+z^4+z^2+z)-\frac{2A}{2+7A}(z^6+z^3+1).
\end{align}
In general,
\begin{align}
    p^{(3)}_{3k-1}(z) &= z^{3k-1} + \frac{A}{2+3(k-1)A-A}\left( \sum_{j=0}^{k-1}z^{3j} + z^{3j+1}  \right) - \frac{2A}{2+3(k-1)A-A}\left(\sum_{j=1}^{k-1} z^{3j-1}\right)\nonumber\\
    &=z^{3k-1} + (-1)^{k-1}\cos\theta^{(3)}_{3k-1}\left( \sum_{j=0}^{k-1}z^{3j} + z^{3j+1}  \right) - 2(-1)^{k-1}\cos\theta^{(3)}_{3k-1}\left(\sum_{j=1}^{k-1} z^{3j-1}\right),\\
    p^{(3)}_{3k}(z)&= z^{3k} + \frac{A}{2+3(k-1)A+A}\left( \sum_{j=0}^{k-1}z^{3j+1} + z^{3j+2} \right) -\frac{2A}{2+3(k-1)A+A}\left( \sum_{j=0}^{k-1}z^{3j} \right)\nonumber\\
    &= z^{3k} + (-1)^{k-1}\frac{\cos\theta^{(3)}_{3k}}{2}\left( \sum_{j=0}^{k-1}z^{3j+1} + z^{3j+2} \right) -(-1)^{k-1}\cos\theta^{(3)}_{3k}\left( \sum_{j=0}^{k-1}z^{3j} \right).
\end{align}
Using \eqref{Eq: monic p} and \eqref{Eq: kappa rho}, the above results lead to \eqref{Eq: P m=3}. Since $m=3,6$ obey the Laplace transformation relations $G_L^{(6)}(z) = G_L^{(3)}(-z)$, similar to $m=1,2$ cases, one can directly conclude the OPUC relation \eqref{Eq: P m=6}.
\end{widetext}

\section{Floquet Sokhotski–Plemelj theorem}
\label{App: I}
We start with the definition of the Floquet
Dirac delta function in frequency space,
\begin{align}
    2\pi \delta_{F}(\omega) = \sum_{n=-\infty}^{\infty} e^{-i \omega n} = 2\pi\sum_{m=-\infty}^{\infty} \delta_D(\omega+2\pi m),
\end{align}
where $\delta_D(\omega)$ is the usual Dirac delta function. The Kronecker delta function in discrete time is related to the difference of  the step functions
\begin{align}
    \delta_{0n} = s(n) - s(n-1),\  s(n) = \Big\{\begin{array}{cc}
     1/2 & \text{if } n \geq 0;\\
     -1/2 & \text{else}. \end{array}
\end{align}
The discrete time Fourier transformation of the above relation is
\begin{align}
    &\sum_{n=-\infty}^{\infty} \delta_{0n} e^{-i\omega n}\nonumber\\
    &= \left( \sum_{n=-\infty}^{\infty} s(n)e^{-i\omega n} \right) - \left( \sum_{n=-\infty}^{\infty} s(n-1)e^{-i\omega n} \right),
\end{align}
and can be further simplified to
\begin{align}
    1 =& \left( \sum_{n=-\infty}^{\infty} s(n)e^{-i\omega n} \right) \nonumber\\
    &- e^{-i\omega}\left( \sum_{n=-\infty}^{\infty} s(n-1)e^{-i\omega (n-1)} \right)\nonumber\\
    =&(1-e^{-i\omega})\left( \sum_{n=-\infty}^{\infty} s(n)e^{-i\omega n} \right).
\end{align}
Finally, we obtain
\begin{align}
    & \sum_{n=-\infty}^{\infty} s(n)e^{-i\omega n}  = \frac{1}{1-e^{-i\omega}} = \frac{1}{2} - \frac{i}{2}\cot\frac{\omega}{2}.
\end{align}
Therefore, we have  
the following relation that will be useful for computing the Laplace transformation
\begin{align}    &\lim_{\epsilon\rightarrow 0^+}\sum_{n=0}^{\infty} e^{-(\epsilon+i\omega) n}\nonumber\\
&= \sum_{n=-\infty}^{\infty} \left(s(n)+\frac{1}{2}\right)e^{ -i\omega n} = \frac{1}{2}  - \frac{i}{2}\cot\frac{\omega}{2} + \pi \delta_F(\omega).
\end{align}
This relation is called the Floquet Sokhotski–Plemelj theorem
\begin{align}
   &\lim_{\epsilon\rightarrow 0^+}\sum_{n=0}^{\infty} e^{-(\epsilon+i\omega) n}\nonumber\\
   &= \lim_{\epsilon\rightarrow 0^+}\frac{1}{1-e^{-\epsilon -i\omega}} = \frac{1}{2} - \frac{i}{2}\cot\frac{\omega}{2} + \pi \delta_F(\omega). 
\end{align}

Below we will apply the this formula to derive the spectral functions for the persistent $m$-period autocorrelations reported in Sec.~\ref{Sec: m-period}. For $m=1$, we first rewrite \eqref{Eq: GL1} such that its expression is ready for Floquet Sokhotski–Plemelj theorem
\begin{align}
    G_L^{(1)}(z) = \frac{z-1+A}{z-1} = (1-A) + \frac{A}{1-z^{-1}}.
\end{align}
According to \eqref{Eq: spectral Laplace}, the spectral function for $m=1$ is
\begin{align}
    \Phi^{(1)}(e^{i\omega}) &= \frac{1}{2\pi}\left(-1 + 2\lim_{\epsilon\rightarrow 0^+} \text{Re}[G^{(1)}_L(e^{\epsilon + i\omega})]\right)\nonumber\\
    &= \frac{1-A}{2\pi} + A \delta_F(\omega).
\end{align}
This derivation proves \eqref{Eq: spectral function m=1}. For generic $m$, $G^{(m)}_L(z)$ is related to $G^{(1)}_L(z)$ as follows
\begin{align}
    G^{(m)}_L(z) =& 1 + \sum_{n=1}^{\infty} A\cos\left( \frac{2\pi n}{m}\right)z^{-n}\nonumber\\
    =& \frac{1}{2}\left( 1 + A\sum_{n=1}^{\infty} (ze^{-i2\pi/m})^{-n} \right)\nonumber\\
    &+ \frac{1}{2}\left( 1 + A\sum_{n=1}^{\infty} (ze^{i2\pi/m})^{-n} \right)\nonumber\\
    =& \frac{1}{2} \left( G^{(1)}_L(ze^{-i2\pi/m}) + G^{(1)}_L(ze^{i2\pi/m}) \right),
\end{align}
where \eqref{Eq: A m-period} is applied in the first line. Therefore, the spectral function $\Phi^{(m)}(e^{i\omega})$ can be related to $\Phi^{(1)}(e^{i\omega})$ as follows
\begin{align}
    \Phi^{(m)}(e^{i\omega}) =& \frac{1}{2\pi}\left(-1 + 2\lim_{\epsilon\rightarrow 0^+} \text{Re}[G^{(m)}_L(e^{\epsilon + i\omega})]\right)\nonumber\\
    =& \frac{1}{2\pi}\left(-1 + \lim_{\epsilon\rightarrow 0^+} \text{Re}\left[G^{(1)}_L(e^{\epsilon + i(\omega-2\pi/m)})\right.\right. \nonumber\\
    &\left.\left. + G^{(1)}_L(e^{\epsilon + i(\omega+2\pi/m)})\right]\right)\nonumber\\
    =& \frac{1}{2} \left[ \Phi^{(1)}(e^{i(\omega+2\pi/m)}) + \Phi^{(1)}(e^{i(\omega-2\pi/m)}) \right].
\end{align}
This result validates the analytic expressions for the spectral function for $m=2,3,4,6$ in Sec.~\ref{Sec: m-period}.

\section{Edge mode eigen-operator of the Majorana chain beyond 2 sublattices}
\label{App: J}
An eigen-operator $\Psi_\lambda$ of the ITFIM satisfies the eigen-operator equation
\begin{align}
    K_I|\Psi_\lambda) = e^{i\omega_\lambda} |\Psi_\lambda),
\end{align}
where $K_I$ has a simple form in the Majorana basis according to \eqref{Eq: Majorana basis matrix element}. Explicitly, the above equation can be expressed as follows
\begin{widetext}
\begin{subequations}
\begin{align}
    e^{i\omega_\lambda}\psi_1 =& \cos\theta_1 \times \psi_1 + \sin\theta_1 \times \psi_2, \label{Eq: psi^m -1}\\
    e^{i\omega_\lambda}\psi_{2j} =& -\sin\theta_{2j-1}\cos\theta_{2j} \times \psi_{2j-1} + \cos\theta_{2j-1}\cos\theta_{2j}\times \psi_{2j} \nonumber\\  
    &+ \sin\theta_{2j}\cos\theta_{2j+1} \times \psi_{2j+1} + \sin\theta_{2j}\sin\theta_{2j+1}\times \psi_{2j+2},\ \forall\ j \geq 1, \label{Eq: psi^m -2}\\
   e^{i\omega_\lambda}\psi_{2j+1} =& \sin\theta_{2j-1}\sin\theta_{2j} \times \psi_{2j-1} - \cos\theta_{2j-1}\sin\theta_{2j}\times \psi_{2j} \nonumber\\  
    &+ \cos\theta_{2j}\cos\theta_{2j+1} \times \psi_{2j+1} + \cos\theta_{2j}\sin\theta_{2j+1}\times \psi_{2j+2},\ \forall\ j \geq 1 \label{Eq: psi^m -3}
\end{align}
\end{subequations}
where we introduce $|\Psi_\lambda) = (\psi_1, \psi_2, \ldots)^\intercal$ in the Majorana basis. By reformulating \eqref{Eq: psi^m -1} and considering the following linear combinations between \eqref{Eq: psi^m -2} and \eqref{Eq: psi^m -3}: addition of \eqref{Eq: psi^m -2} multiplied by $(-\cos\theta_{2j})$ and \eqref{Eq: psi^m -3} multiplied by $\sin\theta_{2j}$, and sum of \eqref{Eq: psi^m -2} multiplied by $\sin\theta_{2j}$ and \eqref{Eq: psi^m -3} multiplied by $\cos\theta_{2j}$, we obtain
\begin{subequations}
\begin{align}
    &\sin\theta_1 \times \psi_2 = \left(e^{i\omega_\lambda}-\cos\theta_1 \right) \psi_1,\\
    &e^{i\omega_\lambda}\sin\theta_{2j}\times \psi_{2j+1} = \sin\theta_{2j-1}\times \psi_{2j-1} + \left( e^{i\omega_\lambda}\cos\theta_{2j} - \cos\theta_{2j-1}\right)\psi_{2j},\ \forall\ j \geq 1, \\
    &\sin\theta_{2j+1}\times \psi_{2j+2} = e^{i\omega_\lambda}\sin\theta_{2j}\times \psi_{2j} + \left( e^{i\omega_\lambda}\cos\theta_{2j} - \cos\theta_{2j+1}\right)\psi_{2j+1},\ \forall\ j \geq 1.
\end{align}
\end{subequations}
The above can be further simplified to
\begin{subequations}
\label{Eq: eigen psi}
\begin{align}
    &\psi_2 = \frac{e^{i\omega_\lambda}-\cos\theta_1}{\sin\theta_1}\psi_1,\\
    &\psi_j = \frac{e^{(-1)^j i \omega_\lambda}\sin\theta_{j-2}}{\sin\theta_{j-1}}\psi_{j-2} + (-1)^j \frac{e^{(-1)^j i \omega_\lambda}\cos\theta_{j-2}-\cos\theta_{j-1}}{\sin\theta_{j-1}}\psi_{j-1},\ \forall\ j \geq 3.
\end{align}
\end{subequations}
However, above equations contain many unspecified parameters, their number being  proportional to the operator space dimension. To reduce the number of independent equations, we have to impose conditions on the coefficients $\{ \psi_j \}$ and Krylov angles $\{ \theta_j \}$. Here, we focus on exponentially localized edge mode solutions with periodic Krylov angles
\begin{align}
    \theta_{lk+j} = \theta_j,\ \forall\, 1 \leq j \leq l,\ k\geq 0,
    \label{Eq: l-sublattice angles}
\end{align}
where $l$ is the period of the Krylov angles. This periodicity leads to a $l$-sublattice structure of the Majorana chain and therefore the eigen-operator obeys this structure as well. In other words, we expect
\begin{align}
    \psi_{lk+j} = c_j \xi^k,\ \forall\,1 \leq j \leq l,\ k\geq 0,
    \label{Eq: l-sublattice psi}
\end{align}
where $\{ c_j \}$ are the coefficients within one sublattice. $\xi$ depicts the localization length of the edge mode and satisfies $|\xi| < 1$. From \eqref{Eq: l-sublattice angles} and \eqref{Eq: l-sublattice psi}, \eqref{Eq: eigen psi} are reduced to $l+1$ equations
\begin{subequations}
\label{Eq: reduced eigen equ}
\begin{align}
    &c_2 = \frac{e^{i\omega_\lambda}-\cos\theta_1}{\sin\theta_1}c_1,\\
    &c_j = \frac{e^{(-1)^j i \omega_\lambda}\sin\theta_{j-2}}{\sin\theta_{j-1}}c_{j-2} + (-1)^j \frac{e^{(-1)^j i \omega_\lambda}\cos\theta_{j-2}-\cos\theta_{j-1}}{\sin\theta_{j-1}}c_{j-1},\ \forall\ 3 \leq j \leq l,\\
    &c_1 \xi = \frac{e^{(-1)^{l+1} i \omega_\lambda}\sin\theta_{l-1}}{\sin\theta_{l}}c_{l-1} + (-1)^{l+1} \frac{e^{(-1)^{l+1} i \omega_\lambda}\cos\theta_{l-1}-\cos\theta_{l}}{\sin\theta_{l}}c_{l},\\
    &c_{2}\xi = \frac{e^{(-1)^{l+2} i \omega_\lambda}\sin\theta_{l}}{\sin\theta_{1}}c_{l} + (-1)^{l+2} \frac{e^{(-1)^{l+2} i \omega_\lambda}\cos\theta_{l}-\cos\theta_{1}}{\sin\theta_{1}}c_{1}\xi.   
\end{align}
\end{subequations}
In order to proceed, one has to know the allowed values of $l$ for a given $\omega_\lambda$. So far we only have empirical ways to explore possible values of $l$. For example, Krylov angles in Fig.~\ref{Fig: Z3} (upper left panel) show 6-periodicity for the $2\pi/3$-mode. Therefore, it is natural to assume $l=6$ for solving $\omega_\lambda = 2\pi/3$ in \eqref{Eq: reduced eigen equ}. Numerically, the Krylov angles can be determined from a given autocorrelation without specifying underlying models, see discussion in \cite{yeh2025moment}. One can first construct an autocorrelation showing edge mode dynamics with frequency $\omega_\lambda$ and then determine $l$ from the patterns in the numerically computed Krylov angles. Below we discuss the details for solving \eqref{Eq: reduced eigen equ} with given $\omega_\lambda$ and $l$.

\begin{figure*}
    \includegraphics[width=0.23\textwidth]{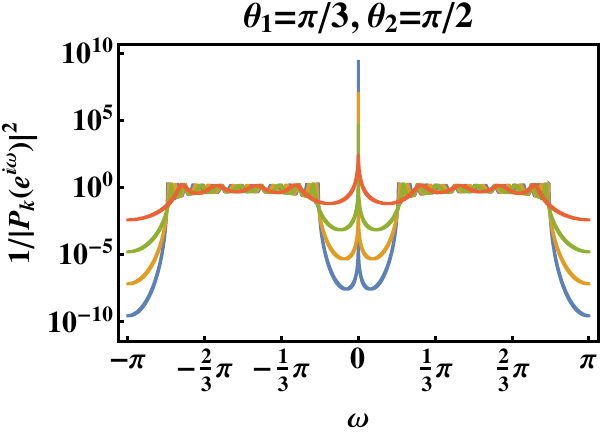}
    \includegraphics[width=0.23\textwidth]{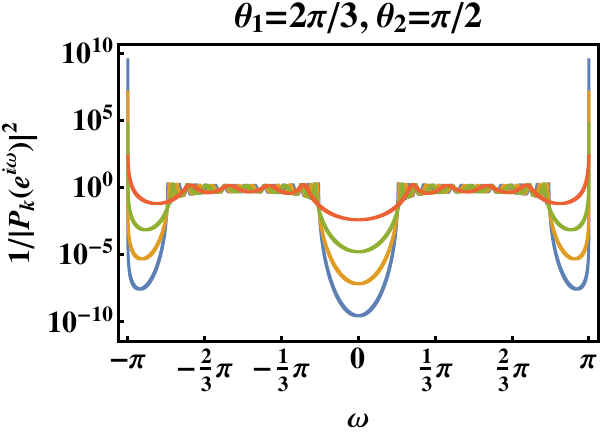}
    \includegraphics[width=0.23\textwidth]{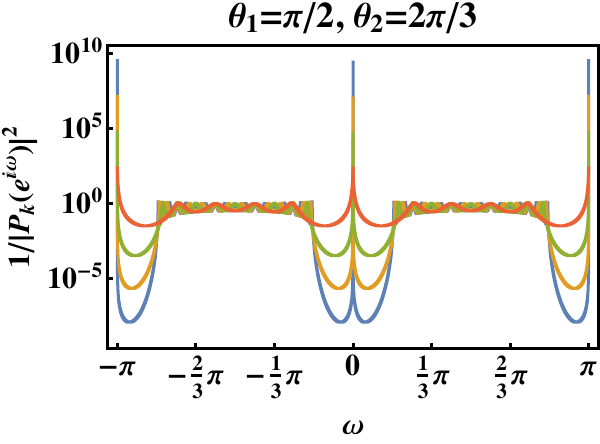}
    \includegraphics[width=0.28\textwidth]{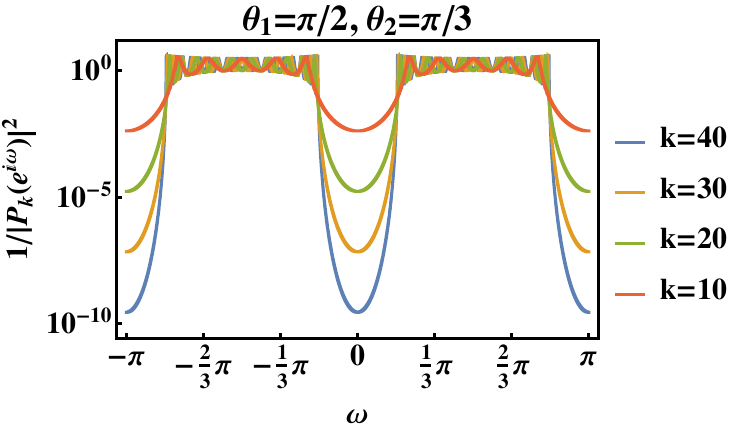}
    \caption{Results for $1/|P_k(e^{i\omega})|^2$ for the Floquet transverse-field Ising model with homogeneous transverse ($\theta_1$) and Ising ($\theta_2$) couplings (green crosses in Fig.~\ref{Fig: Ising Phase Diagram}). The panels from left to right correspond to four different topological phases with the following edge modes: $0$-mode, $\pi$-mode, both $0,\pi$-modes and trivial. The peak positions indicate the edge modes of the corresponding topological phases. The results are consistent with the phase diagram shown in Fig.~\ref{Fig: Ising Phase Diagram}.}
    \label{Fig: Ising}
\end{figure*}

We perform numerically self-consistent computation in \eqref{Eq: reduced eigen equ} to solve for the angles. By setting values for $\omega_\lambda, l, c_1, \theta_1, \theta_2, \ldots, \theta_6$, the first two lines in \eqref{Eq: reduced eigen equ} uniquely determine the values of $c_2, c_3, \ldots, c_l$. Note that one can always set $c_1 = 1$ as linear equations are insensitive to the overall prefactor in the solutions. The last two equations in \eqref{Eq: reduced eigen equ} lead to self-consistent conditions as $\xi$ can be computed separately from each equation. If the initial setting for $\theta_1, \ldots, \theta_l$ are correct, the last two equations in \eqref{Eq: reduced eigen equ} should give consistent results for $\xi$. We focus on $\omega_\lambda = 2\pi/3$ and $l=6$. The self-consistent computation gives three allowed cases for the angles: (i) $\theta_1 = \theta_2 = \theta_4 = \theta_5 = \pi/2$ and $\xi = \tan(\theta_3/2)\cot(\theta_6/2)$, (ii) $\theta_2 = \theta_3 = \theta_5 = \pi/2$, $\theta_1=\theta_4$ and $\xi = \cot(\theta_6/2)$, (iii) $\theta_1 = \theta_3 = \theta_4 = \pi/2$, $\theta_2=\theta_5$ and $\xi = \cot(\theta_6/2)$.

For case (i): $\theta_1 = \theta_2 = \theta_4 = \theta_5 = \pi/2$ and $\xi = \tan(\theta_3/2)\cot(\theta_6/2)$, the analytic solution for the $2\pi/3$-mode is
\begin{align}
    \Psi_{2\pi/3} = \sum_{j=0}^{\infty} 
    &\left[ \cos\left(\frac{\theta_3}{2}\right)\gamma_{1+6j} + e^{i\frac{2\pi}{3}}\cos\left(\frac{\theta_3}{2}\right)\gamma_{2+6j} + e^{-i\frac{2\pi}{3}}\cos\left(\frac{\theta_3}{2}\right)\gamma_{3+6j}\right. \nonumber\\
    & \left.+ e^{-i\frac{2\pi}{3}}\sin\left(\frac{\theta_3}{2}\right)\gamma_{4+6j} + e^{i\frac{2\pi}{3}}\sin\left(\frac{\theta_3}{2}\right)\gamma_{5+6j} + \sin\left(\frac{\theta_3}{2}\right)\gamma_{6+6j}  \right] \frac{\xi^j}{\sqrt{3-3\xi^2}}.
\end{align}
For case (ii): $\theta_2 = \theta_3 = \theta_5 = \pi/2$, $\theta_1=\theta_4$ and $\xi = \cot(\theta_6/2)$, the analytic solution for the $2\pi/3$-mode is
\begin{align}
    \Psi_{2\pi/3} = \sum_{j=0}^{\infty} 
    &\left[ \sin\theta_1\gamma_{1+6j} + \left(e^{i\frac{2\pi}{3}} - \cos\theta_1 \right)\gamma_{2+6j} + \left(e^{-i\frac{2\pi}{3}} - \cos\theta_1 \right)\gamma_{3+6j}\right. \nonumber\\
    & \left.+ \left(e^{-i\frac{2\pi}{3}} - e^{i\frac{2\pi}{3}}\cos\theta_1 \right)\gamma_{4+6j} + e^{i\frac{2\pi}{3}}\sin\theta_1\gamma_{5+6j} + \sin\theta_1\gamma_{6+6j}  \right] \frac{\xi^j}{\sqrt{2+\cos\theta_1}\sqrt{3-3\xi^2}}.
\end{align}
For case (iii): $\theta_1 = \theta_3 = \theta_4 = \pi/2$, $\theta_2=\theta_5$ and $\xi = \cot(\theta_6/2)$, the analytic solution for the $2\pi/3$-mode is
\begin{align}
    \Psi_{2\pi/3} = \sum_{j=0}^{\infty} 
    &\left[ \sin\theta_2\gamma_{1+6j} + e^{i\frac{2\pi}{3}}\sin\theta_2\gamma_{2+6j} + \left(e^{-i\frac{2\pi}{3}} + e^{i\frac{2\pi}{3}}\cos\theta_2 \right)\gamma_{3+6j}\right. \nonumber\\
    & \left.+ \left(e^{-i\frac{2\pi}{3}} +\cos\theta_2 \right)\gamma_{4+6j} + \left(e^{i\frac{2\pi}{3}} +\cos\theta_2 \right)\gamma_{5+6j} + \sin\theta_2\gamma_{6+6j}  \right] \frac{\xi^j}{\sqrt{2-\cos\theta_1}\sqrt{3-3\xi^2}}.
\end{align}
\end{widetext}
The solutions presented above are all normalized according to $\Psi_{2\pi/3}\Psi_{2\pi/3}^\dagger = \mathbb{I}$.

\section{OPUC for the Floquet transverse-field Ising model}
\label{App: K}

\begin{figure}[h!]
    \includegraphics[width=0.32\textwidth]{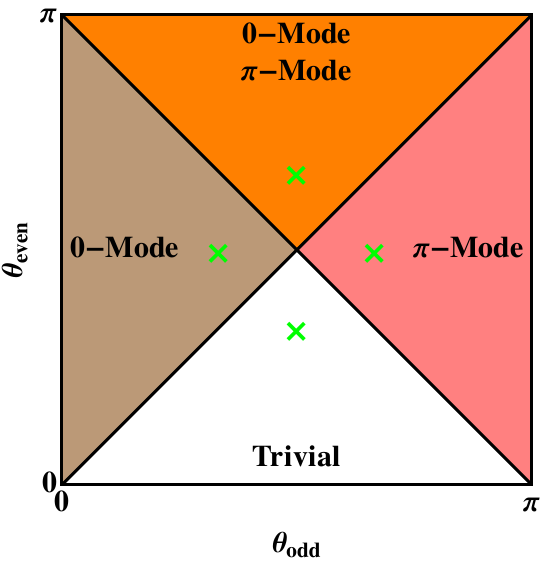}

    \caption{The phase diagram of Floquet transverse-field Ising model with uniform transverse $\theta_{\rm odd}$ and Ising $\theta_{\rm even}$ couplings. The model has four topological phases with distinct edge modes: $0$-mode, $\pi$-mode, both $0,\pi$-modes and trivial. The four pairs of Krylov angles chosen in Fig.~\ref{Fig: Ising} are marked as green crosses. The results in Fig.~\ref{Fig: Ising} are consistent with the phase diagram.}
    \label{Fig: Ising Phase Diagram}
\end{figure}

In this appendix, we present results for OPUC for the conventional Floquet transverse-field Ising model \cite{Sen13,Sondhi16a} in Fig.~\ref{Fig: Ising}. The model has a 2-sublattice structure with the Krylov angles: $\theta_{2k+j}=\theta_j,\ \forall \, 1\leq j \leq 2,\, k\geq 0$. We choose the following four set of Krylov angles, corresponding  to four topological phases: $\theta_{1} = \pi/3,\ \theta_2=\pi/2$ ($0$-mode); $\theta_{1} = 2\pi/3,\ \theta_2=\pi/2$ ($\pi$-mode); $\theta_{1} = \pi/2,\ \theta_2=2\pi/3$ (both $0$ and $\pi$-modes); $\theta_{1} = \pi/2,\ \theta_2=\pi/3$ (trivial). The four choices of Krylov angles are marked as green crosses in the phase diagram of the Floquet transverse-field Ising model, see Fig.~\ref{Fig: Ising}. The peaks in Fig.~\ref{Fig: Ising} indicate the existence of edge-modes and are consistent with Fig.~\ref{Fig: Ising Phase Diagram}. 

The explicit expressions of OPUC, $P_k(\theta_1, \theta_2,z)$, for arbitrary $\theta_1, \theta_2$ have not been reported, to the best of our knowledge. In Refs.~\cite{simon2005orthogonalpart1,ismail2020encyclopedia}, the explicit results are presented for the case $\theta_1 + \theta_2 = \pi$ (where the OPUC are known as the Geronimus Polynomials). We report them here 
\begin{align}
    &P_k(\theta_1, \theta_2 =\pi-\theta_1 ,z) \nonumber\\
    &= \frac{1}{\sin^k\theta_1} \left[ \frac{z - \cos\theta_1}{2^{k-1}}\frac{z_1^k -z_2^k}{z_1 - z_2} - \frac{z\sin^2\theta_1}{2^{k-2}} \frac{z_1^{k-1} -z_2^{k-1}}{z_1 - z_2} \right],
    \label{Eq: Ising Analytic}
\end{align}
where we modify the original expression from Refs.~\cite{simon2005orthogonalpart1,ismail2020encyclopedia} by the Krylov angle parametrization. $z_1, z_2$ is defined as
\begin{subequations}
\begin{align}
    &z_1 = z + 1 + \sqrt{(z-z_+)(z-z_-)},\\
    &z_2 = z + 1 - \sqrt{(z-z_+)(z-z_-)},\\ 
    &z_{\pm} = e^{\pm 2i\arcsin{|\cos\theta_1|}}.
\end{align}
\end{subequations}
We noticed that the analytic expression presented in Ref.~\cite{simon2005orthogonalpart1} has a misprint in that powers of $2$ in the denominator of \eqref{Eq: Ising Analytic} are missing. 

According to Sec. \ref{Sec: m-period}, we can generalize the above results to two additional cases: (i) $\theta_1 = \theta_2 = \theta$ by using the relation between 1 and 2-period cases \eqref{Eq: m=2 Krylov angle} and \eqref{Eq: P m=2}, and (ii) $\theta_1 = \pi/2$ and $\theta_2 = \theta$ by  using the relation between 1 and 4-period Krylov angles \eqref{Eq: m=4 Krylov angle} and \eqref{Eq: P m=4}. 
In particular, according to \eqref{Eq: m=2 Krylov angle} and \eqref{Eq: P m=2} , we have the following relations
\begin{align}
    &P_k(\pi-\theta_1, \theta_2 = \pi-\theta_1,z)\nonumber\\
    &= (-1)^kP_k(\theta_1, \theta_2 = \pi-\theta_1,-z).
\end{align}
By relabeling $\pi-\theta_1 \rightarrow \theta$, we obtain the results for $\theta_1 = \theta_2 = \theta$ case as below
\begin{align}
    P_k(\theta, \theta,z) = (-1)^kP_k(\pi-\theta,\theta,-z ).
\end{align}
Next, by \eqref{Eq: m=4 Krylov angle} and \eqref{Eq: P m=4}, we obtain OPUC for $\theta_1 = \pi/2$ and $\theta_2 =\theta$ as follows
\begin{subequations}
\begin{align}
    &P_{2k}(\pi/2, \theta, z) = (-1)^kP_{k}(\theta,\theta,-z^2) = P_k(\pi-\theta,\theta, z^2),\\
    &P_{2k+1}(\pi/2, \theta, z) = zP_{2k}(\pi/2, \theta, z).
\end{align}
\end{subequations}
At the dual unitary point, \eqref{Eq: Ising Analytic} is simplified by the following relation
\begin{align}
    \frac{z_1^k-z_2^k}{2^{k-1}(z_1-z_2)} = \frac{(2z)^k-2^k}{2^{k-1}(2z-2)} =  \sum_{j=0}^{k-1}z^j,\ \forall\ \theta_1 = \theta_2 = \frac{\pi}{2}.
\end{align}
Then one recovers the dual unitary result: $P_k(z) = z^k$.


%

\end{document}